**Title: Insular intracranial activity identifies multiple facial expressions via diverse, intermixed temporal patterns at the single-contact level**

Yingyu Huang[1#], Lisen Sui[2#], Liying Zhan [1,3], Chaolun Wang[1], Zhihan Guo[1], Yanjuan Li[2], Xiang Wu[1*]

(#These authors contributed equally to this work)

[1]Department of Psychology, Sun Yat-Sen University, Guangzhou, China

[2]Department of Neurology, The Second Affiliated hospital of Guangzhou University of Chinese Medicine (Guangdong Provincial Hospital of Chinese Medicine), Guangzhou, China

[3]School of Education, Zhaoqing University, Zhaoqing, Guangdong, China

* Correspondence:

Xiang Wu

Department of Psychology, Sun Yat-Sen University, 132 Waihuan East Road, Higher Education Mega Center, Guangzhou, Guangdong, China, 510006

E: wuxiang3@mail.sysu.edu.cn

# Graphic Abstract

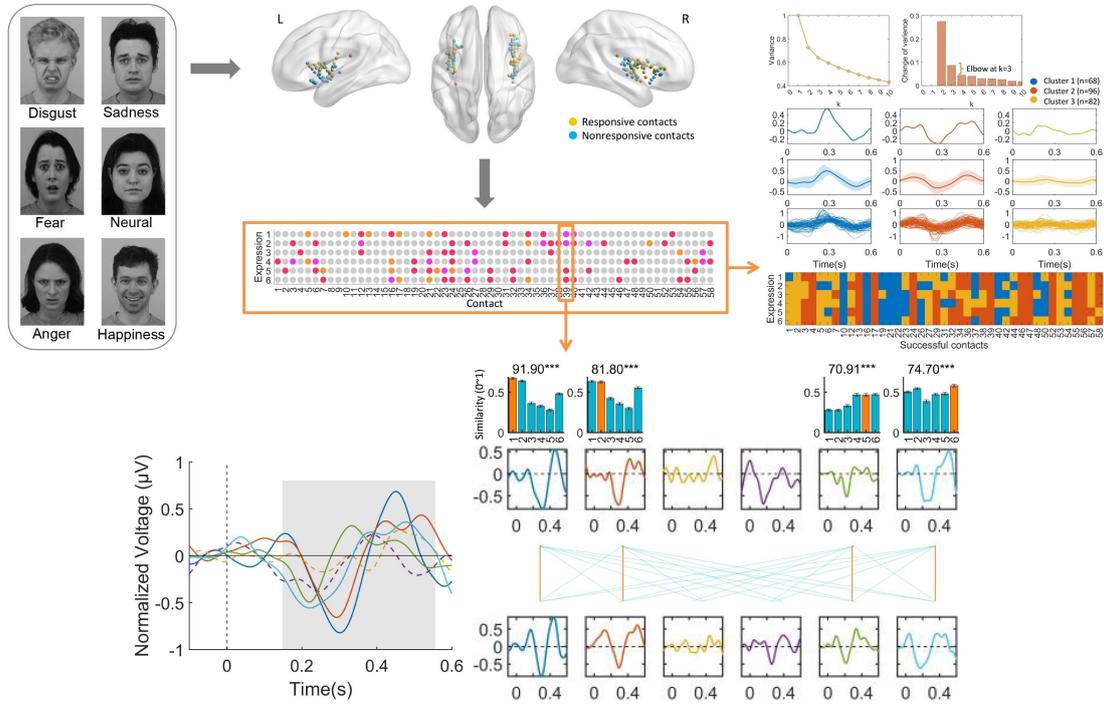

# Abstract

How neural representations in the insular cortex support emotional processing remains poorly understood, and the extent to which the insula is specialized for disgust processing remains debated. We recorded stereoelectroencephalography data from the insula while human subjects with implanted electrode contacts performed a facial emotion recognition task involving disgusted, fearful, angry, sad, neutral, and happy expressions. Expression category specificity of insular activity was assessed via pairwise comparisons of within- and between-category pattern similarities, capturing both the shape and scale of event-related potentials (ERPs) and event-related spectral perturbations (ERSPs; theta to high-gamma frequency ranges). Insular activity successfully identified all investigated expressions, mediated by diverse ERP responses intermixed across the insula. In contrast to the marked heterogeneity of insula ERP responses, the fusiform face area exhibited convergent ERP responses across expressions and contacts, with ERSPs also contributing substantially to expression identification. These findings not only elucidate the insula's neural mechanisms underlying facial emotion perception, but also establish a potential single-contact-level neural substrate for how the insula leverages its heterogeneous response profiles to act as a key hub for versatile cognitive and emotional functions.

## Introduction

The insular cortex, as the primary gustatory cortex, is vital for survival by enabling avoidance of harmful food [1,2]. The disgust-related processing of the insula is involved not only in primary sensory processing but also in emotional processing in broader cognitive functions, making it a key brain hub for integrating sensory, emotional, and cognitive systems [3–8].

Despite the generally acknowledged specific role of the insula in disgust processing, conflicting evidence and different viewpoints exist. For example, in a functional magnetic resonance imaging (fMRI) study, Schienle et al. [9] showed that the insula is also involved in processing non-disgust facial emotions and that disgust processing also occurs in other areas, argued against the insula's specificity for disgust processing. A meta-analysis of fMRI results similarly fails to support the insula's specificity for disgust processing [10]. Using electrical stimulation, previous research further found that stimulation of the insula induces fear and anxiety responses [11].

Disgust processing in the human insula has primarily been investigated via neurophysiological and non-invasive brain imaging approaches. Clinical behavioral observations [12,13] are limited by small sample sizes and individual lesion variability. Non-invasive brain imaging methods including fMRI and scalp electroencephalography (EEG) / magnetoencephalography (MEG) [7,14–17] have poor temporal and spatial resolution.

Intracranial EEG (iEEG) via implanted electrodes provides a valuable approach for observing neural electrical activity with high spatial and temporal precision underlying cognitive functions [18–21]. To explore the insula's role in disgust processing, Krolak-Salmon et al. [22] used stereoelectroencephalography (sEEG) and observed that, in electrode contacts within the ventral anterior insula, event-related potential (ERP) components about 300 ms after stimulus onset were larger for viewing disgust faces compared to faces with other expression types (happy, fearful, and neutral). Interestingly, such disgust-specific insular activity was not observed in an auditory emotion recognition study (including disgust, fear, anger, sadness, neutral, and happiness), which investigated event-related spectral perturbation (ERSP) activity in the high-gamma band but found no differences in insular activity across emotion types [23].

Thus, the insula is a key brain region for emotional experience, and disgust is a fundamental emotion critical for survival. While the insula has been ascribed a specialized role in disgust processing, the underlying temporal, spatial, and spectral features of the insular activity, and the extent to which the insula is specific to disgust processing, remain major issues of debate. The present study was designed to address the issues by leveraging intracranial EEG, which provides high spatial and temporal resolution recording of insular activity [18]; and similarity-based pattern analysis, which allows sensitive quantification of variations in insular activity [24,25].

sEEG was used to record insular cortical activity while human subjects with implanted electrodes in the insula (Fig. 1b) performed a facial expression recognition task involving six expression categories: disgust, fear, anger, sadness, neutral, and happiness [26] (Fig. 1a). The expression category specificity of insular activity was analyzed by pairwise comparisons between within-category and between-category similarities of post-stimulus timecourse patterns [24,25] (Fig. 1c). The neural activities investigated included the ERPs and the ERSPs from theta to high-gamma frequency bands [21]. The present study aimed to address two key questions: (1) What are the temporal, spatial, and spectral characteristics of insular activity that underlie facial expression identification? (2) To what extent is the insula specialized for disgust processing?

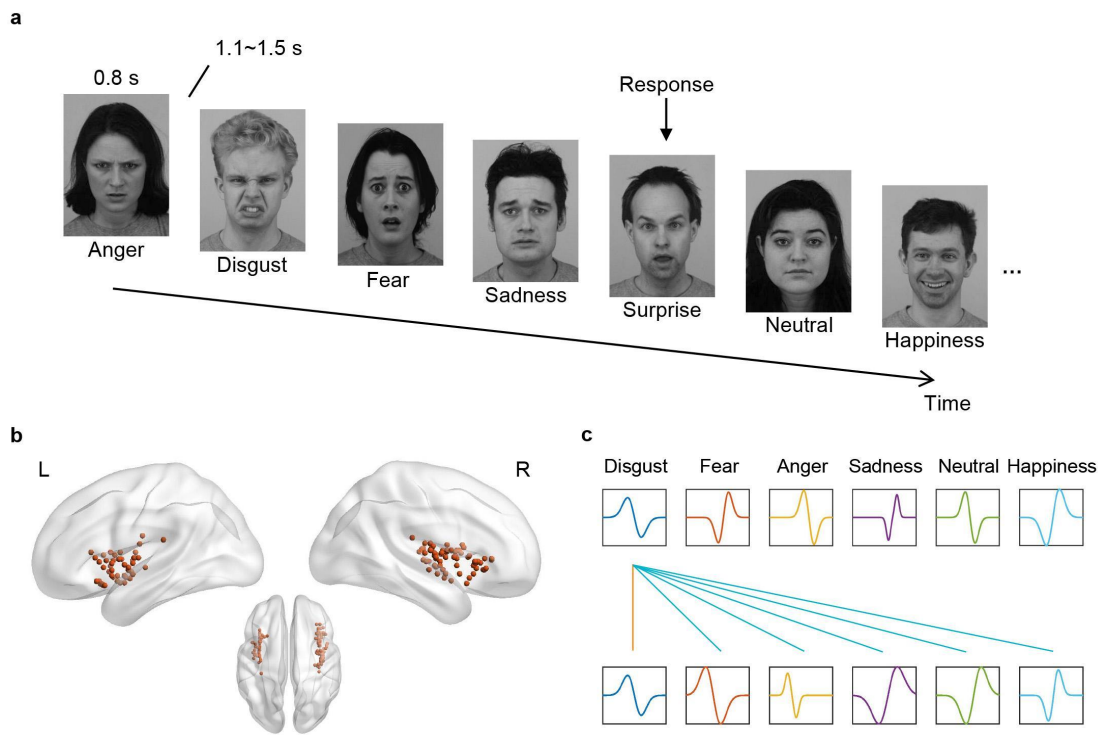

**Fig. 1. Illustration of the study design**. **a**. Depicts the experimental stimuli and procedure. Subjects were instructed to detect surprise faces among disgust, fearful, angry, sad, neutral, and happy faces. (Target surprise expression was excluded from neural activity analysis). **b.** Displays the insular bipolar electrode contacts (*n*=122, left hemisphere 46, right hemisphere 76) overlaid on a surface brain template (detailed contact information is provided in Table S1). **c**. Schematic diagram illustrating identification of an expression via temporal patterns of ERPs or ERSPs at a single electrode contact. Within-category similarity referred to the similarity between the patterns of odd- and even-numbered trials for that expression (orange line); between-category similarity referred to the similarity between the pattern of odd-numbered trials for that expression and the pattern of even-numbered trials for another expression (blue line). Within-category similarity was compared to each of the between-category similarities, and the accuracy of identifying the expression was regarded as the proportion of comparisons where within-category similarity exceeded between-category similarity. The similarity was estimated using Euclidean distance-based distance, which captures both the shape and scale of event-related response patterns. The event-related responses shown in the diagram are simulated by biphasic Gaussian waveforms. The present study adopted the similarity-based

approach given its simplicity, high efficiency, and particularly favorable interpretability (Haxby, 2012; Grootswagers et al., 2017).

## Results

### Behavioral performance

The mean±SD *d'* of the behavioral responses to the surprise faces was 1.92±0.66 (Wilcoxon signed-rank test against *d'*=1, *p*=.001), and the mean±SD response time was 593.51±55.56 ms. The performance was consistent with previous reports [27], indicating that the subjects understood the task and complied with the task instruction. Two subjects with *d'*<1 were excluded from subsequent brain imaging data analyses.

### Expression identification by temporal patterns of insular activity

We carried out pattern analysis for the ERPs and the ERSPs of theta, alpha, beta, gamma, and high-gamma frequency bands.

### ERPs

### Evaluation of responsive contacts and important time periods

Prior to pattern analyses, procedures were performed to identify contacts with valid evoked neural responses and to assess the importance of temporal features for differentiating between facial expressions.

To evaluate responsive contacts, for each contact, the trials of expression conditions were averaged and a Wilcoxon signed-rank test was performed to compare the value of each time point in the post-stimulus duration (0-600 ms) against the mean baseline value (-100-0 ms). A contact was indicated as responsive if at least one response peak was identified, measured as 10 (i.e., 20 ms length) consecutive significant (*p*<.001) time points. The evaluation procedure identified 58 responsive insular contacts (Fig. S1a). Only the responsive contacts were subjected to following pattern analyses. To evaluate important time periods, for the post-stimulus timecourses at each contact, principal component analysis (PCA) was performed on a trial-averaged matrix with dimensionality T x C (where T referred to the number of time points and C referred to the number of expressions), with time points treated as features. PCs that explained 80% of the variance were extracted, and the importance of original features (i.e., the time points) was assessed via loadings of the extracted PCs. At least 10 (i.e., 20 ms length) consecutive important time points were considered. Important time periods were parametrically assessed according to the top 75%, top 50%, or top 25% of total importance (Fig. S1c). Subsequent pattern analyses were based on the important time periods evaluated according to the top 50% of total importance, which captured the dominant temporal pattern variance across expressions (see Fig. S7 for pattern analysis based on the full post-stimulus time duration, which yielded consistent results).

**Pattern analysis**

Successful identification was achieved for all the investigated facial expressions (Fig. 2a). Specifically, disgust, fearful, angry, sad, neutral, and happy faces were identified at 14, 17, 9, 16, 15, and 15 contacts, with corresponding mean±SD accuracies of 69.5±11.4, 72.4±11.2, 69.8±6.7, 72.6±11.9, 69.9±10.4, and 69.7±8.4, respectively. Overall, 39 among the 58 responsive contacts showed successful expression identification, yielding a mean±SD accuracy of 70.8±10.4 across all expressions (Fig. 2b).

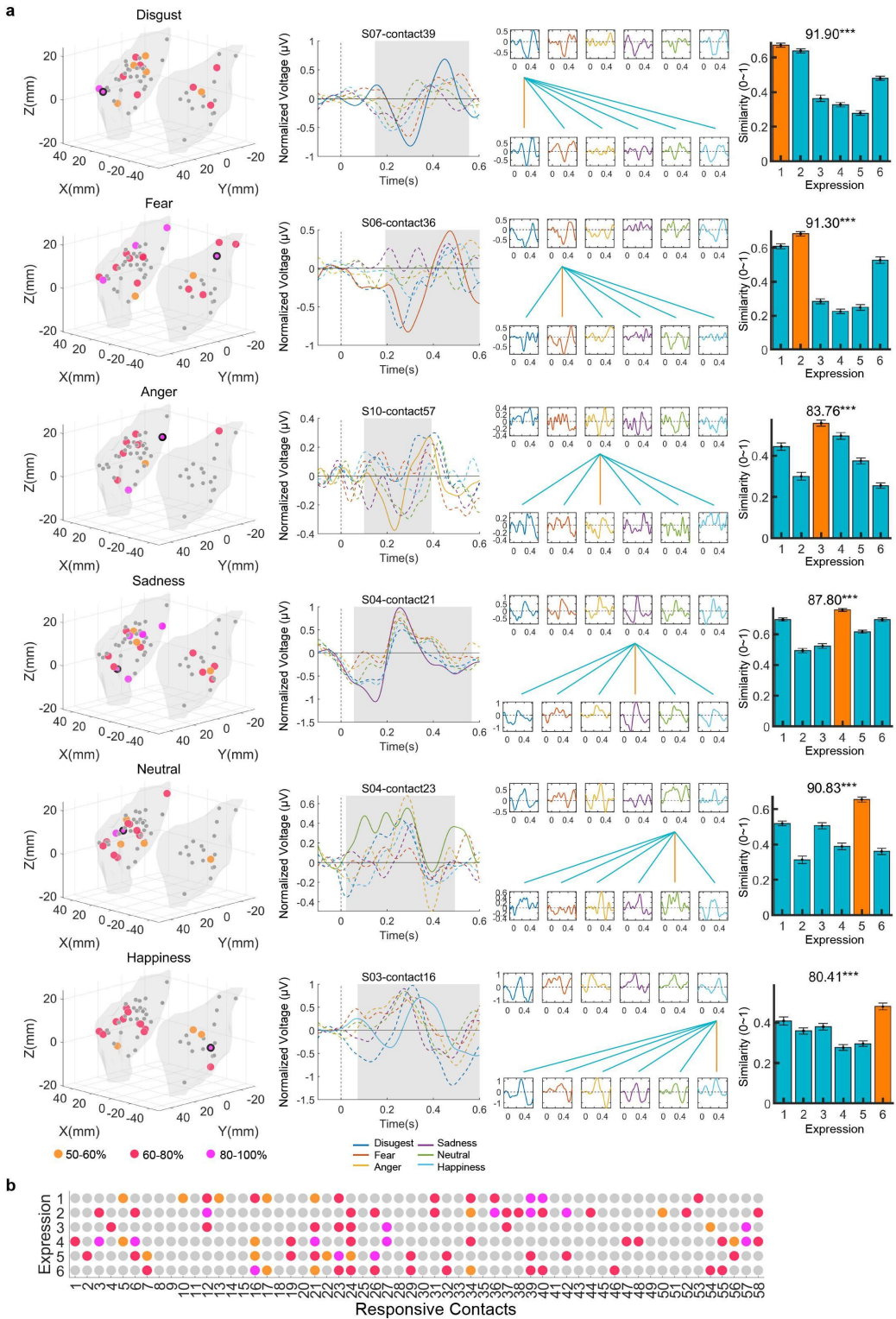

**Fig. 2. Successful facial expression identification by insular ERP patterns.** **a**. Each row shows identification results for an expression. Left panel: Contacts where the expression was successfully identified (random permutation test, $p_{corrected}$ < .05) are presented in a three-dimensional space (viewed from the left anterior aspect of the brain), where MNI coordinates are indicated. Left and right insular contours are indicated. Identification accuracies are color-indicated for ranges 50-60%, 60-80%, and 80-100% (responsive contacts without successful identification of the expression are marked in gray). Right panel: Depicts, at a representative contact (circled in the left panel), how the expression was identified. Left subpanel: ERP timecourses to all the expressions. Middle subpanel: Schematic of the patten analysis method using Euclidean distance-based similarity (see Fig. 1c). (For illustration, waveforms are from one of 200 permutations, see Methods). Gray shaded areas indicate the important time periods according to top 50% of total importance. Subject IDs and Contact IDs are annotated; Contact IDs correspond to responsive contacts only. Right subpanel: Within- category and between-category similarity values for the expression. Error bars indicate 95% CIs. Accuracy is labeled; Statistical significance: one asterisk indicates $.01 \leq p_{corrected} < .05$; two asterisks indicate $.001 \leq p_{corrected} .01$; three asterisks indicate $p_{corrected} < .001$. Expressions 1-6 denote disgust, fear, anger, sadness, neural, and happiness, respectively. (Within- and between-category similarity values for all successful identifications are provided in Fig. S2). **b**. Matrix representation of successful expression identifications, with rows indicating the six expressions and columns indicating the responsive contacts. Other conventions are as in Fig. 1.

**Spatial characteristics**

Visual inspection (Fig. 2a) indicated that contacts with successful expression identification were diffusely distributed across the insula and were intermixed with contacts without successful identification (despite the left insula having more total number of contacts (Fig. 1a) and responsive contacts (Fig. S1a)); which was statistically verified (Kruskal-Wallis H test: $p>.05$ for x, y, and z coordinates). Fig. 3a-b illustrate the spatial distributions of contacts that identified different expressions, suggesting that the contacts identifying distinct expressions were intermixed at the single-contact level (Kruskal-Wallis H test: $p>.05$ for x, y and z coordinates).

Of note, in insular parcellation schemes the ventral anterior insula (VAI), dorsal anterior insula (DAI), and posterior insula (PI) are the commonly studied insular subregions; among them, the VAI has been proposed to be more involved in emotional processing [7,17,28]. Fig. 3c maps the contacts with successful identification according to the VAI, DAI, and PI parcellation. While the small number of VAI successful contacts (3) (Fig. 3d) precluded detailed analysis of VAI data and balanced cross-subregion comparisons, it was clear that expression identification was not restricted to the VAI but was also distributed across the DAI and PI.

Moreover, for the current observation of the diffuse and intermixed expression identification across the insula, it was worth further noting that individual contacts typically identified more than one expression (Fig. 2b): each contact was capable of identifying 1 to 5 expressions (of the 41 contacts with successful identification, only 14 identified a single expression). This surprising observation provided further strong evidence against specialized processing of distinct expressions at spatially segregated insular subareas. The ability of single contacts to identify multiple expressions would be attributed to the discriminative information contained in timecourse variations across different expressions, which was further explored below.

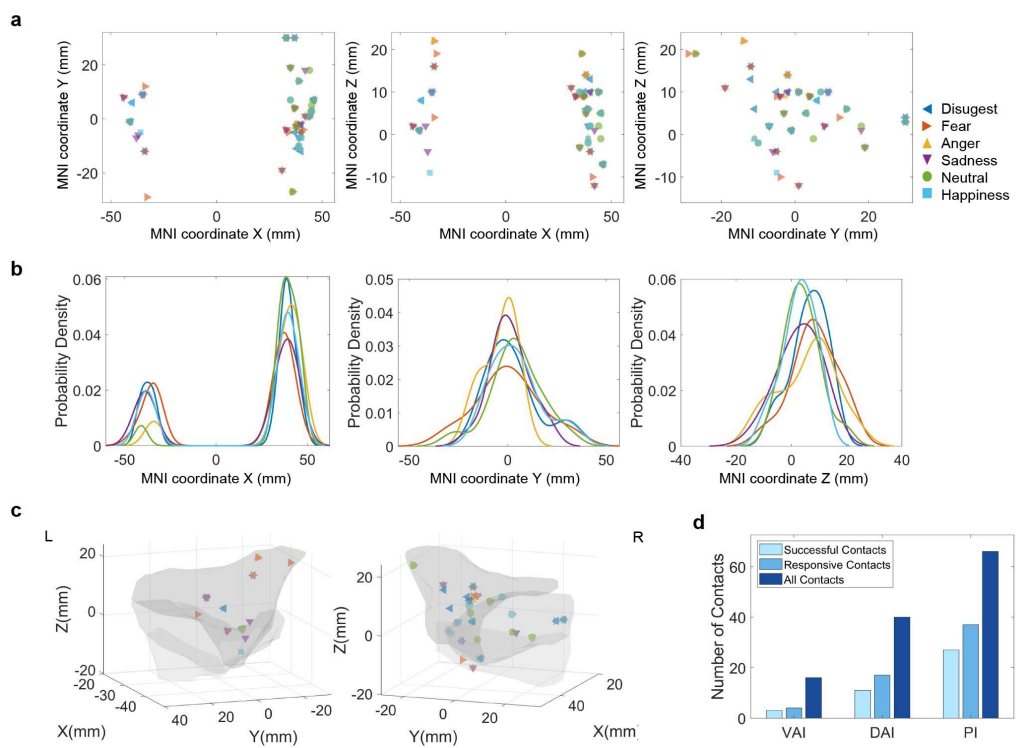

**Fig. 3. Spatial characteristics of insular activity in expression identification. a**. Contacts identifying different expressions are pooled and projected onto the y-z, x-y, and x-z planes, respectively. **b**. Distributions of contacts successfully identifying the six expressions (probability density functions of a Gaussian mixture model), along the x, y, z coordinates, respectively. **c**. Presents the contacts with successful identification in a three-dimensional space, with the three insular parcellated subregions annotated (ventral anterior insula, VAI; dorsal anterior insula, DAI; and posterior insula, PI). **d**. Number of contacts with successful identification, alongside responsive contacts and all contacts, within each of the three subregions. Other conventions are as in Fig. 2.

**Temporal characteristics**

Consider successful contact 39 as a representative example, where four expressions (disgust, fear, neutral, and happiness) were identified (Fig. 4a). Identification of an expression depended on the specific temporal pattern of the timecourse to that expression. When an expression's temporal pattern (e.g., disgust or fear) was more specific to that expression and less similar to patterns of other expressions, it allowed the expression to be distinguished from more other expressions, resulting in higher identification accuracy. Conversely, when an expression's temporal pattern (e.g., neural or happy) was less specific to that expression and more similar to patterns of other expressions, the expression could only be distinguished from fewer other expressions, corresponding to lower accuracy.

In addition, as shown above (Fig. 2; Fig. S2), a given expression could be identified at different contacts by distinct temporal patterns (Fig 4b further illustrates a representative case). Thus, the diverse temporal patterns associated with different expressions across widely distributed insular contacts provided abundant discriminative information for facial expression identification – information that was essential to elucidate how, and to what extent, an expression could be identified at the single-contact level.

To further characterize the diversity of temporal patterns across expressions and insular contacts, we carried out a cluster analysis using the k-medoids clustering algorithm [20], in which ERP timecourses of all expressions from all contacts with successful identification were included. The results confirmed the heterogeneity of insular temporal patterns and further revealed two key attributes of this diversity. (1) About two thirds of insular timecourses exhibited similar response profiles, as those in Clusters 1 and 2; the other third did not show consistent response profiles, as those in Cluster 3 (Fig. 4d). The response profile in Cluster 1 showed a positive deflection spanning approximately 200-400 ms and peaking at approximately 300 ms, while the response profile in Cluster 2 showed a negative deflection over a similar time range. (2) Response profiles were intermixed across expressions and insular contacts: among the 41 contacts with successful identification, just 4 displayed the same Cluster 1 profile for all expressions, and only 10 showed the same Cluster 2 profile for all expressions (Fig. 4f). This high heterogeneity of insular temporal patterns was markedly different from the convergent temporal profiles observed in the fusiform face area (FFA; see Fig. 6-7 in the following FFA analysis).

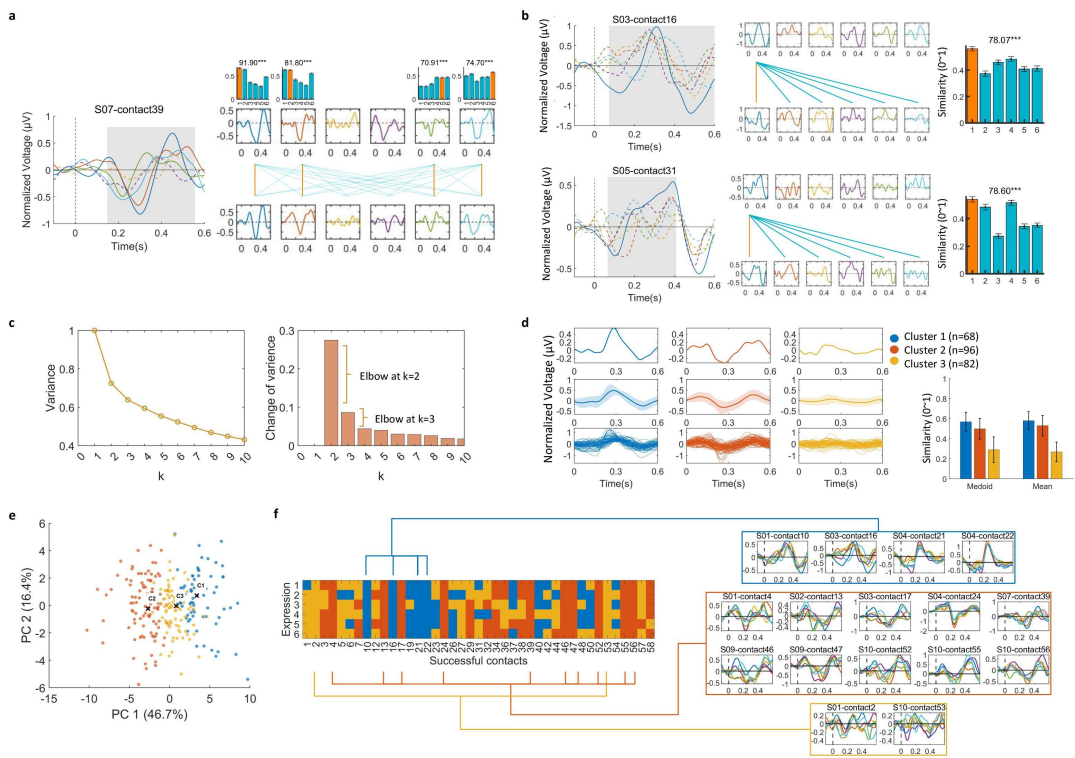

**Fig. 4. Temporal characteristics of insular activity in expression identification. a**. At Contact 39, in addition to disgust (Fig. 2a, top row), fear, neutral, and happy expressions were also identified. The temporal pattern to fear was similar to that to disgust, whereas the patterns to neutral and happiness were different from those to disgust and fear. Within-category and between- category similarity values for an expression are displayed above its timecourses to better illustrate multiple expression identifications at a single contact. **b**. Two additional contacts are presented to further illustrate how disgust could be identified at different contacts by distinct temporal patterns (complementing the example from Contact 39). **c**. k-medoids clustering with the elbow method was applied to ERP timecourses from all expressions at all contacts with successful identification (totally 6 x 41 = 246 timecourses). Left: Normalized variance (sum of electrode to cluster center distances, normalized to the sum at k=1) as a function of cluster number (k). Right: Change in variance as a function of k. Discernible drops in variance change occurred after k=2 and k=3, indicating possible elbows (point of transition from steep to moderate slope of variance). (Because clustering requires equal-length data, the clustering analysis used the same post-stimulus time period for all timecourses). **d**. Shows results for k = 3. For the three clusters (left to right), the top row depicts ERP timecourses of the medoids (center timecourses that best represent cluster timecourses); the middle row depicts mean ERP timecourses (shaded regions around the mean indicate SDs); and the bottom row depicts individual timecourses. Additionally, average similarity values of individual timecourses to the medoid and the mean timecourses are shown for each cluster. (k=2 results were presented in Fig. S8a). Cluster 1 response profile exhibited a positive deflection over ~ 200-400 ms, while Cluster 2 response profile showed a wide negative deflection over the same approximate time range. In Cluster 3, there were substantial variability in the patterns of timecourses; neither the medoid nor the mean timecourse could identify a consistent profile across individual timecourses. **e**. Data (each point represents a timecourse) are visualized on their first two principal components (medoids marked with 'x'). **f**. Distributions of clusters across expressions and contacts. Of 41 successful contacts, only 16 contacts exhibited identical response

profiles for all expressions: 4, 10, 2 for Clusters 1-3, respectively (individual timecourses from these contacts are further indicated). Other conventions are as in Fig. 1-2.

## ERSPs

### Evaluation of responsive contacts and important time periods

The procedures evaluating responsive contacts and important time periods were conducted for the ERSP of each frequency band. A total of 2, 2, 9, 2, and 1 responsive insular contacts were identified for the theta, alpha, beta, gamma, and high-gamma bands, respectively (Fig. S3). Relative to ERPs, far fewer contacts passed the evaluation procedure for ERSPs.

### Pattern analysis

Successful identification was only found for the theta, alpha, and high-gamma frequency bands, and within each of these band only a few expressions could be identified (Fig. 5). Overall, in comparison to ERPs, substantially less information in ERSPs were utilized in identifying expression. The limited number of contacts with successful expression identification would not allow further adequate estimation of their spatial distribution.

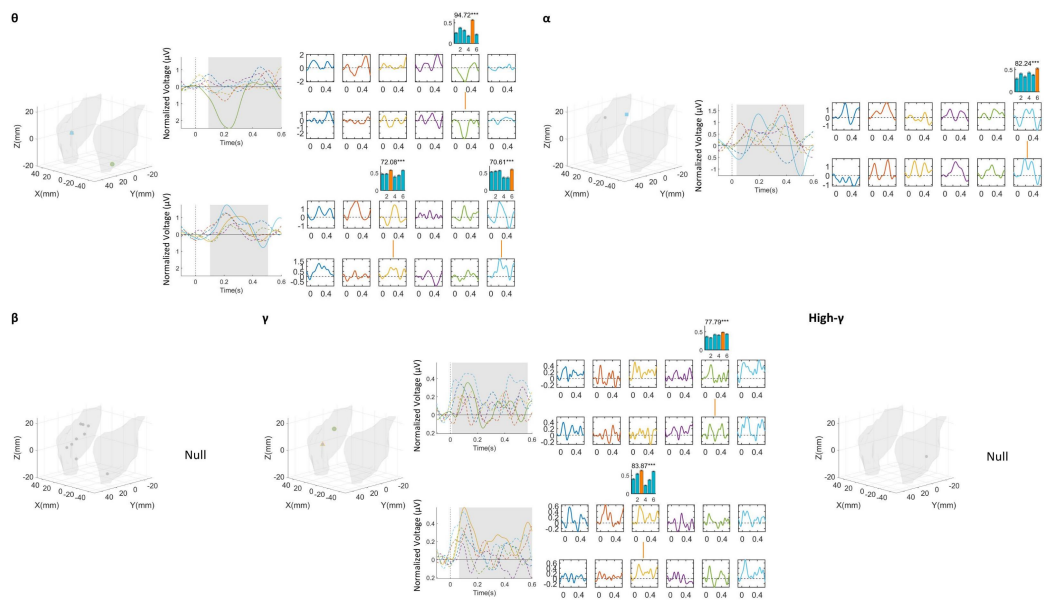

**Fig. 5. Successful facial expression identification by insular ERSP patterns.** Shows the ERSP results for each frequency band; bands with no successful identification are labeled as Null. Left panel: Contacts with successful identification are presented in a three-dimensional space. Different expressions are marked by distinct maker shapes. Right panel: Depicts expression identification for all successful contacts. Other conventions are as in Fig. 2 and 4.

### Expression identification by temporal patterns of FFA activity

The fusiform face area on the ventral visual pathway is specialized for face perception and has been found to be engaged in facial expression recognition [29–32]. Two subjects in the current study also had electrodes implanted in the FFA (Sbj06 had 8 left FFA contacts; Sbj09 had 2 left and 6 right FFA contacts; totally 10 left and 8 right FFA contacts). Analysis of FFA activity provided an extensively characterized reference to better understand insular responses during expression recognition.

### ERPs

The procedure evaluating responsive contacts revealed 15 responsive contacts among the 18 FFA contacts (Fig. S4). (Further statistical analysis of contact spatial distribution was not conducted due to the limited number of FFA contacts).

Successful identification was found for all investigated facial expressions (Fig. 6a). Specifically, disgust, fearful, angry, sad, neutral, and happy faces were identified at 4, 8, 5, 4, 7, and 6 contacts, respectively, with corresponding mean±SD accuracies of 65.3±11.2, 76.0±12.2, 69.8±9.0, 68.2±10.3, 68.5±5.0, and 64.5±7.0. Overall, 14 of the 15 responsive contacts showed successful expression identification, yielding a mean±SD accuracy of 69.3±9.6 across all the expressions (Fig. 6b).

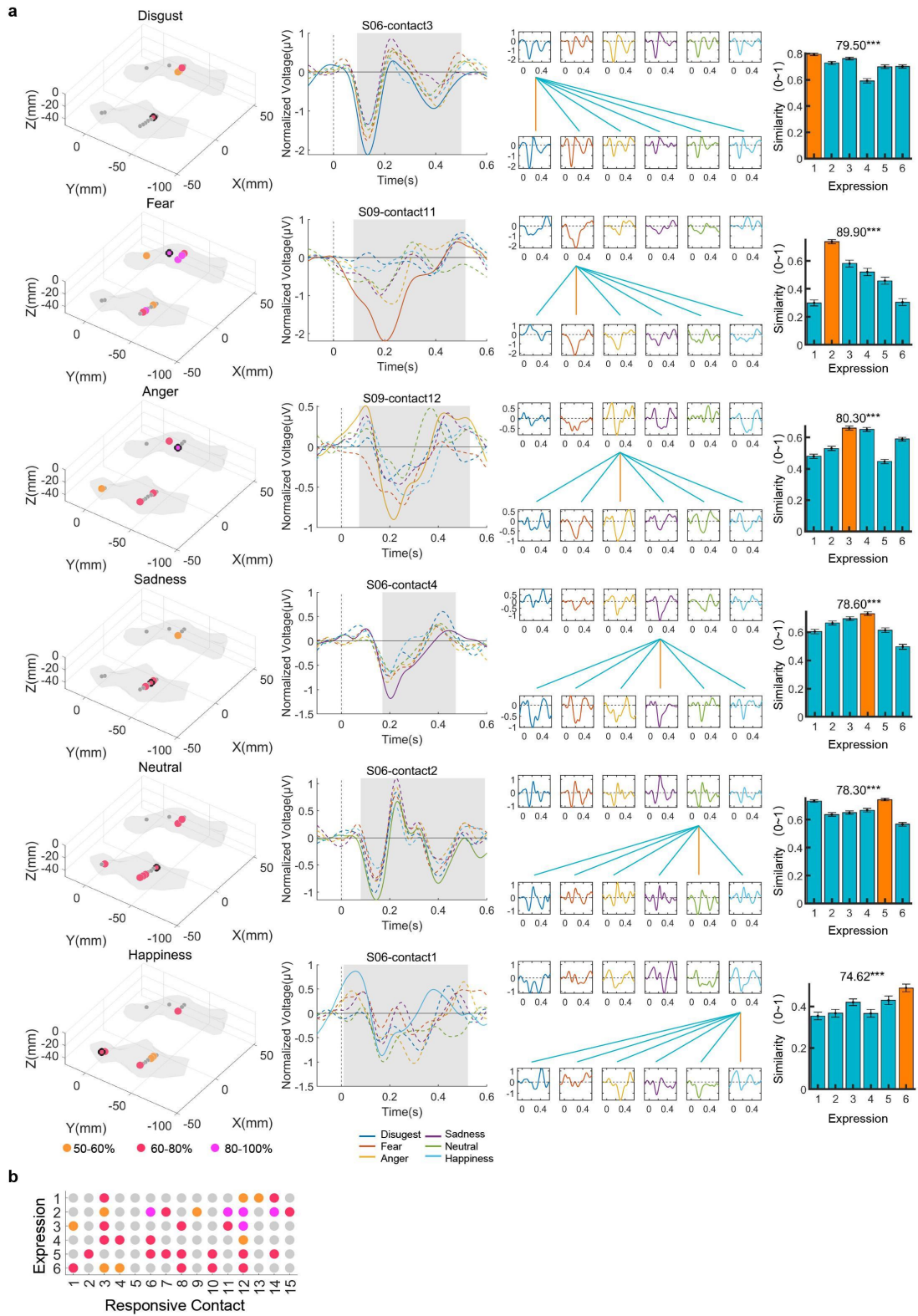

**Fig. 6. Successful facial expression identification by FFA ERP patterns**. **a**. Each row shows the identification results for an expression. Left panel: Contacts with successful identification are presented in a three-dimensional space (viewed from the left posterior aspect of the brain). Right panel: Depicts expression identification for a representative contact (Fig. S5 shows the within-category and between-category similarity values for all FFA contacts with successful identification). **b**. Matrix representation of successful expression identifications. Other conventions are as in Fig. 2.

Similarly to the insular results, multiple expressions could be identified at a single FFA contact and an expression could be identified at distinct contacts (Fig. 6b). However, unlike the high diversity of insular response patterns (Fig. 4c-f), clustering analysis revealed that FFA response patterns were far less diverse across expressions and contacts (Fig. 7). (1) All FFA timecourses exhibited similar response profiles, specifically in the four divided clusters (Fig. 7b). Particularly, response profiles in Cluster 3 and Cluster 4 featured an early negative deflection (peaked at ~ 170 ms for Cluster 3 and ~ 200 ms for Cluster4), consistent with the visual evoked potential (VEP) typically observed in the FFA (e.g., in previous sEEG studies (Krolak-Salmon et al. 2003)). (2) Responses profiles were distributed across expressions and contacts in a more convergent manner: of the 14 contacts with successful identifications, over half displayed identical response profile for all expressions (Fig. 7d).

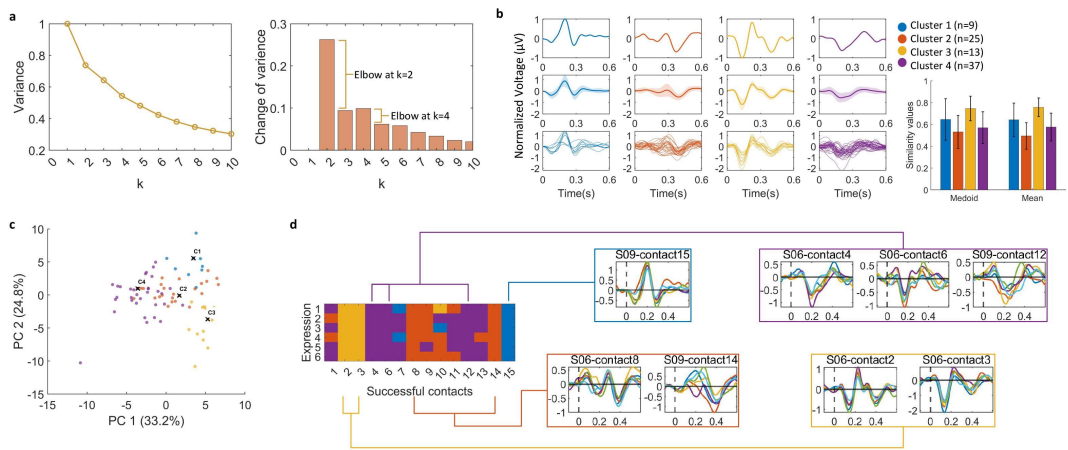

**Fig. 7. Temporal characteristics of FFA activity in expression identification. a**. k-medoids clustering with the elbow method was applied to ERP timecourses of all expressions at all contacts with successful identification (totally 6 x 14 = 84 timecourses). Discernible drops in variance change were observed after k=2 and k=4,

indicating potential elbows. **b**. Shows the results for k=4. Cluster timecourses exhibited similar response profile for all clusters: an early positive deflection peaking at ~200 ms for Cluster 1; a later negative deflection at ~400 ms for Cluster 2; an early negative deflection peaking at ~170 ms for Cluster 3, and an early negative deflection peaking at ~200 ms for Cluster 4. Notably, the response profiles of Cluster 3 and 4 are consistent with the typical VEPs (see Fig. 8a for further peak analysis of the VEPs). (See Fig. S8c for k=2 results). **c**. Data are visualized on their first two principal components. **d**. Distributions of clusters across expressions and contacts. Of the 14 contacts with successful identifications, 8 exhibited identical response profiles for all expressions: 1, 2, 2, 3 for Clusters 1-4, respectively; their timecourses well complied to their cluster medoid and mean time courses. Other conventions are as in Fig. 4 and 6.

**ERSPs**

The procedure evaluating responsive contacts revealed 4, 4, 8, 5, and 5 responsive FFA contacts for the theta, alpha, beta, gamma, and high-gamma bands, respectively (Fig. S6).

Successful identification was found for all frequency bands, though not all expressions were identified within each band (Fig. 8). Thus, different from the insula where limited ERSP information was engaged in expression identification (Fig. 5), the FFA utilized substantially more ERSP information.

Temporal profiles varied across frequency bands (e.g., narrower peaks for higher frequencies, especially for high-gamma responses) and were not further explored via clustering analysis, given the limited number of contacts with successful identification per frequency band.

θ

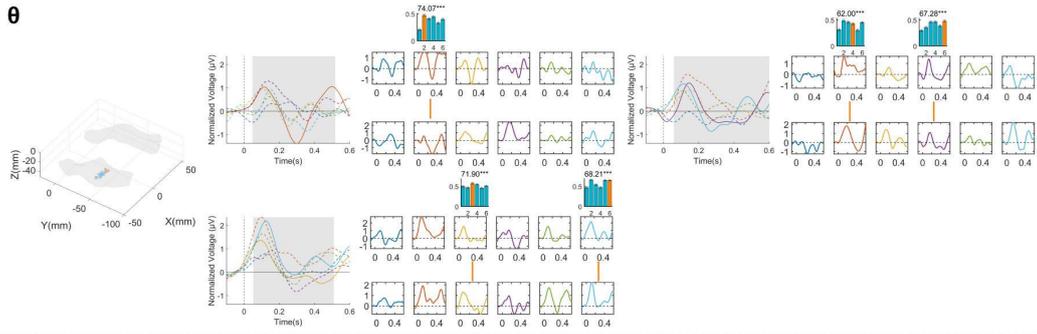

α

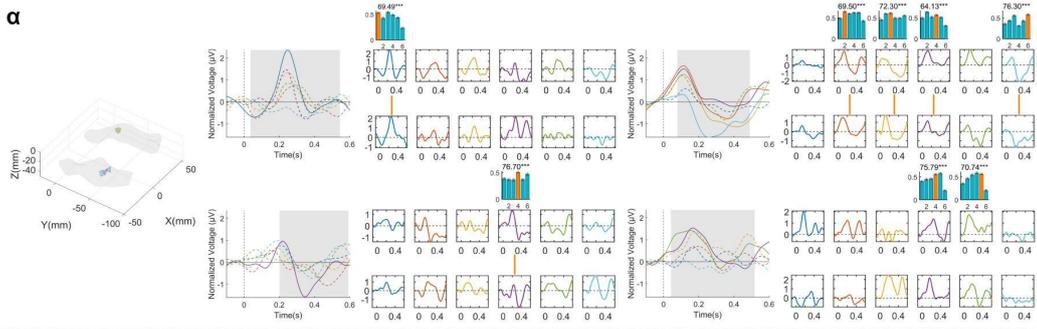

β

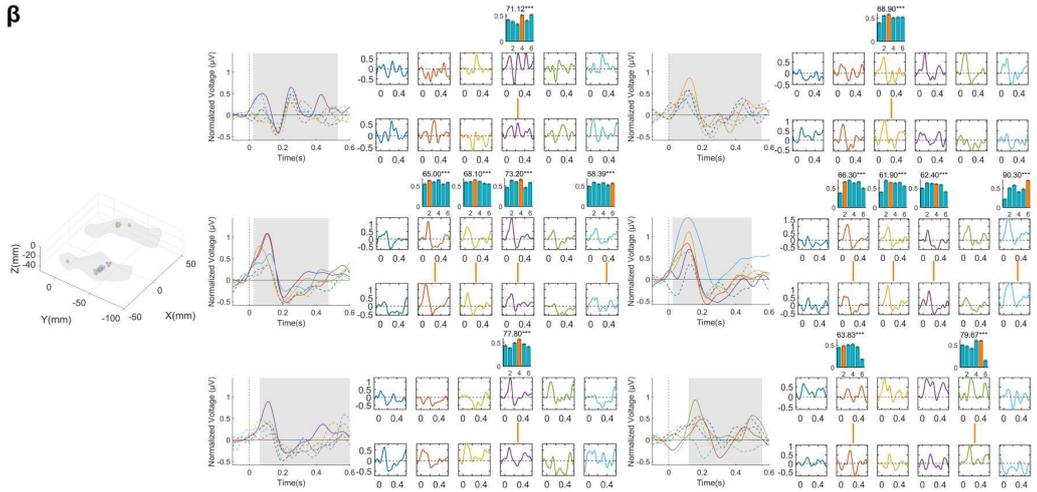

γ

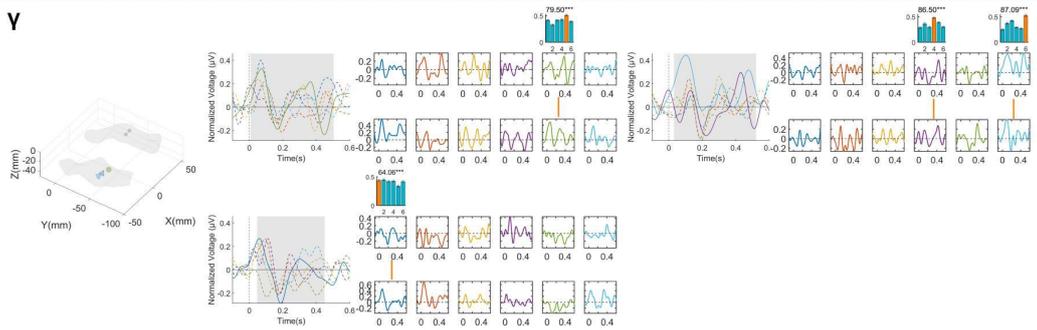

High-γ

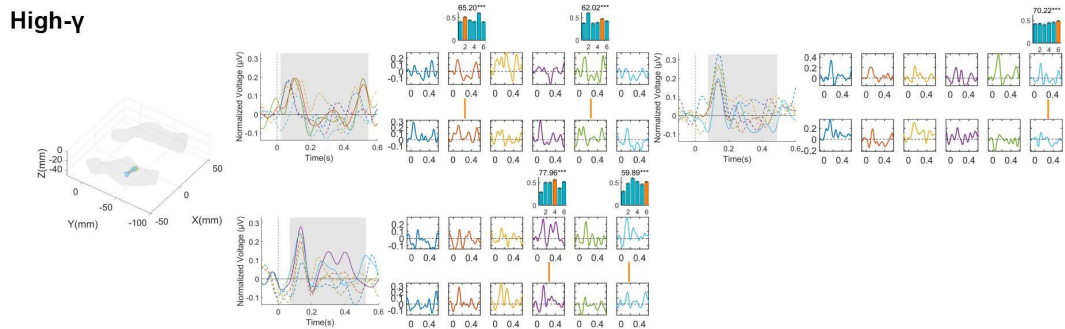

**Fig. 8. Successful facial expression identification by FFA ERSP patterns**. Each row shows ERSP identification results for a frequency band. Left panel: Contacts with successfully identification are presented in a three-dimensional space. Different expressions are marked by distinct maker shapes. Right panel: Illustration of expression identification at all successful contacts. Other conventions are as in Fig. 4-6.

**Sensitivity of Euclidean distance-based pattern analysis in identifying event-related responses to multiple facial expressions**

To discriminate between event-related responses (ERPs and ERSPs) to distinct expressions, the present study employed Euclidean distance-based pattern analysis. Here, we further validate the sensitivity of this method, relative to that of the univariate peak analysis and the correlation distance-based pattern analysis.

**Euclidean distance-based pattern analysis vs. univariate peak analysis**

We showed that expressions were identified via VEPs in the FFA (Fig. 7) – a finding that initially appears to contradict a previous ieeg study reporting null FFA VEP result for expression discrimination (Krolak-Salmon et al., 2003). That study analyzed FFA VEP peak amplitudes and latencies but found no significant differences between expressions. For a given event-related response waveform, traditional peak analysis focuses on evaluating limited time ranges with large amplitudes. In contrast, multivariate pattern analysis emphasizes that the response waveform could carry richer information in its pattern composed of the variation of large and small amplitude data points. We therefore suggest that univariate peak analysis may lack sufficient sensitivity to detect the subtle differences embedded in neural waveforms to distinct expressions. While the current pattern analysis demonstrated successful expression identifications (Fig. 7), Fig. 9a shows that VEP peak analysis did not possess this high level of sensitivity.

**Pattern analysis based on Euclidean distance vs. correlation distance**

Similarity between neuroimaging data patterns is most commonly assessed using correlation distance. The correlation-based method primarily quantifies the shape but not the scale of the pattern, and is better suited for capturing patterns over long data sequences. However, for capturing the patterns of the short event-related responses (as examined in the present study), we suggest that the scale of the pattern is also essential. Fig. 9b presents a toy model demonstrating how the correlation method can completely overlook the scale of one cycle of a biphasic Gaussian or sine waveform (which mimics the positive and negative deflections typically appeared in event-related responses).

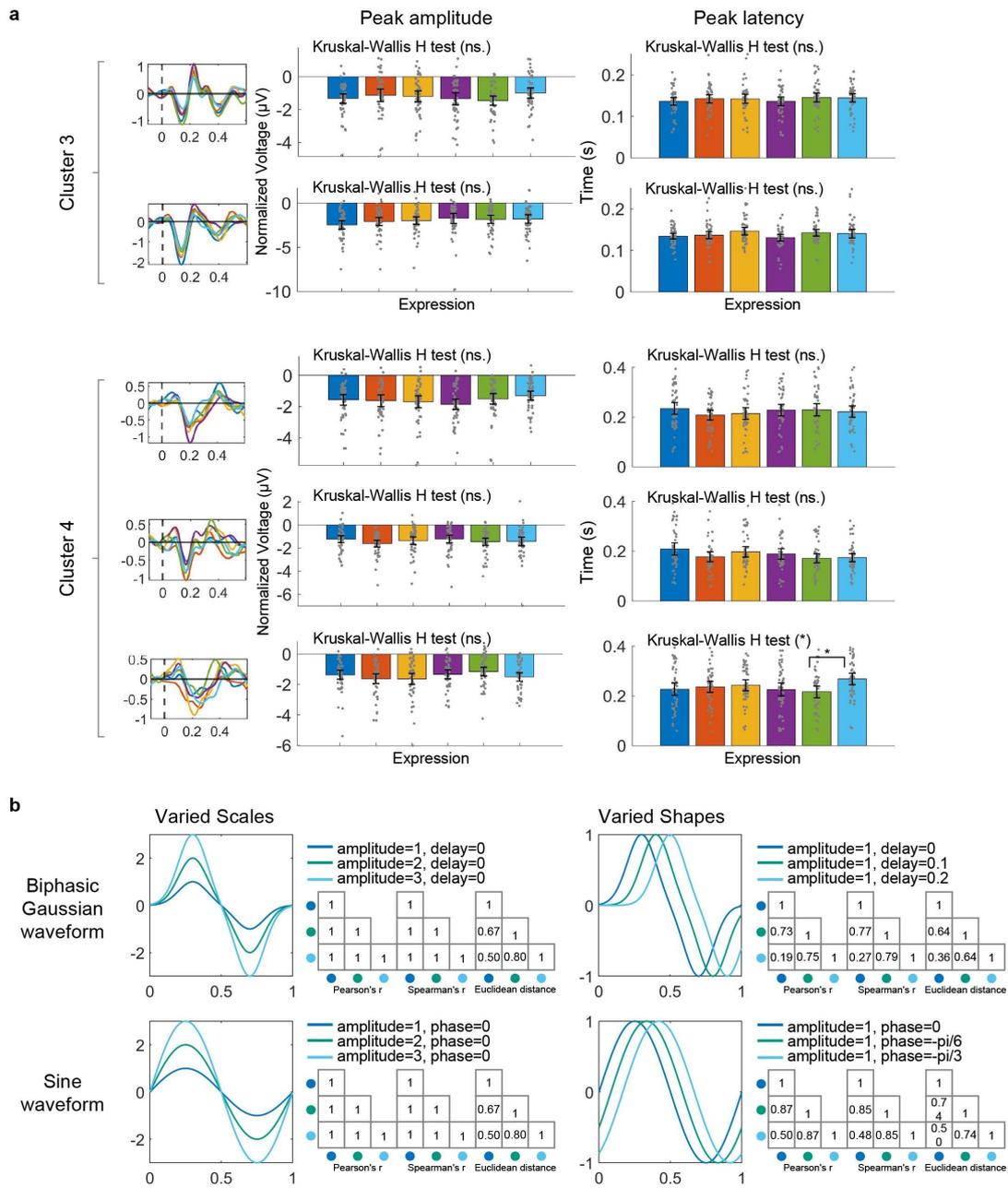

**Fig. 9. Illustration of the sensitivity of Euclidean distance-based pattern analysis in capturing event-related response patterns**. **a**. Peak analysis results of VEPs at FFA contacts. FFA contacts exhibiting identical VEP profiles across all expressions were from Clusters 3 and 4 in the clustering analysis (Fig. 7d) (VEPs waveforms are reproduced as in insets for convenience). Peak amplitudes (assessed over a 20-ms window centered at the peaks) and peak latencies were extracted from single trials for VEP peaks (denoted by arrows) in response to the six expressions. For each contact, the mean±95% CIs of peak amplitudes (left) and latencies (right) are plotted for the

six expressions, respectively. Small dots represent data from individual trials. The Kruskal-Wallis H test was performed to assess differences across depressions; post-hoc tests were conducted with Wilcoxon signed-rank tests if the Kruskal-Wallis H test revealed significant effects across expressions. "ns" indicates non-significant; one, two, and three asterisks indicate $.01 \leq p < .05$, $.001 \leq p < .01$, and $p < 0.001$, respectively. Significant differences across expressions were observed only for peak latency at one contact. **b**. Simulates one cycle of a biphasic Gaussian (top) or sine waveform (bottom) representing a positive deflection and a negative deflection in event-related responses; with manipulated scale defined by amplitude (left) and shape defined by delay for biphasic Gaussian or phase for sine (right). Whereas Euclidean distance-based similarity method captured all these variations, correlation-based similarity method (both Pearson and Spearman correlation) completely ignored the amplitude modulations (similarity values are displayed in the lower triangular similarity matrix ).

## Discussion

The neural mechanisms underlying insular emotional processing remain unclear, and the extent to which the insular cortex is specialized for disgust processing has long been a topic of debate [1,2,9,11]. The present intracranial EEG study revealed how the insular activity leverages its rich temporal, spatial, and spectral features to enable the identification of multiple facial expressions.

### Spectral characteristics

The ERPs and the ERSPs are the primary event-related electrophysiological activities. ERPs are broadband response (though historically and in practice, they are often low-pass filtered with high-frequency components treated as noises). They are phase-locked to stimulus event and are considered as post-synaptic activity. ERSPs are frequency-band-specific response that are non-phase locked and thought to reflect ongoing neuronal oscillations [33,34].

In the present results, successful expression identifications in the insula were primarily found for ERP data. This aligns broadly with prior insular SEEG studies that reported expression-associated differences in ERPs (Krolak-Salmon et al., 2003) but not high-gamma ERSPs [23].

Notably, unlike the insula, for the FFA the capability of expression identification was found for both ERP and ERSP data. While FFA involvement in facial expression recognition has been investigated with modalities including fMRI and EEG, systematic investigation of ERSPs across multiple frequency bands as performed in the present study, remains rare [29–32]. The current results not only highlight a distinction between the insula and FFA in their utilization of neural oscillations for expression decoding, but also indicate the contribution of neural oscillations across a wide range of frequency bands within the FFA to expression recognition.

**Spatial characteristics**

Located deep within the lateral sulcus, the insula exhibits structural and functional heterogeneity; the VAI, DAI, PI are the most commonly studied parcellated insular subregions and have been linked to a wide range of functions [7,17,28]. It has been suggested that the VAI is more connected to the limbic system and is involved in autonomic functions and emotional processing; the DAI has stronger connections with prefrontal regions and is associated with cognitive and executive functions; and the PI primarily integrates sensory and motor signals and is engaged in sensorimotor functions [15–17].

The present results showed a diffuse and intermixed distribution of the insular contacts exhibiting successful expression identification. While the finding might appear to contradict existing knowledge of specialized insular subregions for distinct functions, we suggest that the high spatial resolution of sEEG allowed the present study to advance understanding of how representation of different emotion categories are distributed in the insula. Similarly, an intracranial EEG study investigating the insula in episodic memory found that "these sites were sparsely located across the insular cortex and not concentrated in any specific subregion of the insula" [35]. Such sparse and intermixed distributions have also been observed in intracranial EEG studies of other functional brain areas, including language networks [20] and motor cortices [36].

**Temporal characteristics**

In the present results, all FFA responses exhibited similar profiles, with responses profiles largely convergently distributed across expressions and contacts --- an observation particularly evident for the VEPs.

Notably distinct from the FFA --- an inferior temporal region specialized for facial processing, insular responses in the present study displayed high heterogeneity: approximately one-third of the responses did not show consistent patterns, and insula response profiles were distributed in an intermixed manner across expressions and contacts. This marked diversity of insular temporal patterns aligns with the insula's role as a key hub of the cognitive and emotional systems. Considering the versatile cognitive and emotional functions the insula is involved in, the heterogeneous response profiles at the single-contact level may represent a neural substrate that enables the insula's engagement and collaboration with other brain areas across a broader range of emotional tasks. In contrast to lower-level sensory cortices --- where responses are more stereotyped and modality-dependent (e.g., the VEPs), the insula may tune its responses in a more diverse manner to integrate and coordinate signals across sensory, emotional, and cognitive systems in broader task contexts. Additionally, the deflections in insular response profiles (peaking at ~300 ms) occurred later than those in FFA response profiles (peaking at ~170-200 ms), consistent with previous observations (Krolak-Salmon et al., 2003). Crucially, our finding that only a small subset of insular contacts showed identical profiles across expressions further indicates that the insula operates as a flexible hub --- one that leverages localized response variability to mediate diverse emotional behaviors (Chapman and Anderson, 2012; Gogolla, 2017; Benarroch, 2019).

In addition, the high diversity of insular response patterns also presented a methodological difficulty for predefining interested time windows for individual contacts. Instead of relying on subjective predefinition --- whether based on limited prior knowledge or visual inspection (infeasible given the response diversity), we employed a data-driven PCA-based method to evaluate important time periods for individual contact. It is further worth mentioning that direct utilization of the post-stimulus duration (i.e., without the selection of temporal features) yielded consistent results, albeit with slightly lower overall accuracy (see "Expression identification by insular ERP patterns of the post-stimulus time duration" in the supplementary material). This result is unsurprising as pattern analysis is less dependent on specific time ranges. In scenarios where computational efficiency is a priority, the step of predefining individual time ranges might be non-essential (Metzger et al., 2022; Willett et al., 2021).

**Regarding specialized insular processing of disgust**

The present findings do not favor the notion of specialized insular processing for disgust. However, import cognitive and methodological considerations require to be taken into account to avoid potential overstatements or biased interpretations.

First, the present task investigated basic facial expressions at a perceptual level. On the one hand, as the primary gustatory cortex, the insula is specialized for sensory disgust processing [1]. On the other hand, given its extensive connections with other brain structures, the insula has been shown to contribute to a diverse range of higher-order emotional and cognitive functions [4,5,7]. The cognitive level of processing and the specific aspects of the emotional task should be taken into account when interpreting insular emotion processing. Relatedly, higher-level emotional processing may involve more broadly distributed brain areas and networks. The insula serves as a key hub in the emotional network, connecting distinct systems subserving sensory, emotional, and cognitive processing [37]. The present work focused on analyzing activity within the insula, and the results do not rule out the possibility of disgust-specificity in interactions between the insula and other emotional and cognitive regions.

Second, a critical distinction between correlational and causal approaches in neuroimaging research should be emphasized. For example, activity patterns in ventral object-processing regions can successfully predict the viewing of faces versus other object categories—even when the FFA is excluded—yet this finding does not contradict the face recognition impairments caused by FFA lesions [24]. It remains an open question how lesions in insular subareas may induce selective impairments in the recognition of specific face expressions.

**limitations**

The present task employed eight facial identities, and future investigations should incorporate more face identities to enhance generalizability. The number of insular contacts varied substantially across subjects, and the distribution of contacts between the left and right insula was unbalanced at both the individual and group levels. Additionally, only a small number of contacts were localized in the VAI. Further studies with a larger sample of insular contacts --- particularly those more evenly distributed across insular subareas, are needed for a more comprehensive characterization of the spatial distributions of distinct emotional processing mechanisms within the insula.

**Conclusion**

The neural mechanisms underlying emotional processing in the insula remain poorly understood. The present study utilized the high spatial and temporal resolution of sEEG recordings and the sensitivity of pattern analysis method to analyze ERPs and ERSPs elicited by facial expressions. The results demonstrated that insular activity could reliably identify all the investigated expressions, mediated by the diverse ERP patterns intermixed across the insula.

These results highlight the capability of insular activity to decode perceived facial expressions, supported by the high heterogeneity of event-related response pattern at the single-contact level. Not only do the findings elucidate the neural mechanisms underlying facial expression perception, but they also lay a neural foundation for the insula's role as a core hub in emotional and cognitive systems. Future studies should build on this work by incorporating single-contact insular activity pattern analysis with a broader range of emotional task paradigms to further explore the diverse, intermixed insular response profiles in emotional processing.

# Methods

### Subjects

Thirteen patient subjects (7 males; aged 14 to 28 years, mean ± SD = 21.2 ± 5.9) undergoing a presurgical evaluation of drug-resistant epilepsy in Guangdong Provincial Hospital of Chinese Medicine participated in this study. All subjects had normal or corrected-to-normal vision. The research protocols in this study followed the tenets of the Declaration of Helsinki and were approved by the Ethics Committee of Guangdong Provincial Hospital of Chinese Medicine and the Institutional Review Board of Psychology Department of Sun Yat-Sen University. All participants provided written informed consent before their participation.

### Data acquisition

### Behavioral task

The subjects sat in front of a Lenovo Yangtian V340-14-IIL laptop computer with a monitor (60-Hz refresh rate, 1920 x 1080 resolution, and 31.0 cm x 18.5 cm) and a Cherry MX Board 6.0 G80-3930 computer keyboard. The viewing distance was 50 cm. 56 grayscale emotional face images (4 males and 4 females; 7 expressions:

disgust, fear, anger, sadness, surprise, neutral, and happiness) were used. Stimuli were presented for 0.8 s in a random order, with an inter-stimulus interval randomized between 1 ~ 1.5 seconds (Fig. 1a). The face images were taken from the Karolinska Directed Emotional Faces (KDEF) stimulus set of pictures of facial affect [26]. The images were repeatedly presented 6 times in 3 blocks, resulting in 48 trials per expression type. Surprise faces served as the target; subjects were required to press a key using their right index finger upon detecting a target to ensure that subjects kept attentive during the experiment. Due to the distinct target detection and response mechanisms involved, surprise face trials were not included in further EEG analyses [25,38]. The stimuli subtended a visual angle of 2° x 2.7°, presented at the screen center on a black background. A red fixation point (0.2° visual angle in diameter) was presented continuously at the screen center. Stimuli were presented using Psychtoolbox3 for Matlab running on Ubuntu 18.04.

**SEEG data recording**

As part of the presurgical evaluation, depth electrodes were stereotactically implanted and the SEEG was continuously monitored. The electrodes were 0.8 mm in diameter and had 8 to 16 contacts, and the contacts were 2mm in length and 1.5 mm in separation. The SEEG was recorded using the Nihon Kohden system digitized at 2000 Hz. Electrode placement was according to clinical needs only.

**Data analysis**

SEEG data were analyzed offline using a custom processing routine involving MATLAB (for basic signal and statistical processing), SPM 12 (for structural MRI and CT data processing), xjView, MRIcron and BrainNet Viewer (for MRI image visualization), EEGLAB (for EEG data pre-processing), and mfeeg (for basic EEG signal processing).

**Electrode contact localization**

The MRI-CT fusion images were obtained by superposition of preoperative T1 MR

images and postoperative computed tomography (CT) images, using the procedure implemented in SPM12. Normalized Montreal Neurological Institute (MNI) coordinates of electrode contacts were obtained after transforming the MRI-CT fusion image to a MNI template. Anatomical labeling of contacts that belonged to the insula was performed according to the Automated Anatomical Labeling (AAL) atlas, and parcellation of insular subregions (DAI, VAI, PI) was based on the scheme in Deen et al. [16].

**sEEG data preprocessing**

The EEG was downsampled to 500 Hz, 0.5 Hz high-pass filtered and notch-filtered at 50, 100, 150, 200, and 250 Hz. The bipolar reference method was applied, whereby the signal of each contact was subtracted from that of its internal neighbor contact on the same electrode, yielding a virtual contact positioned at the midpoint between the two original contacts [19,39]. Artifact contaminated electrode contacts were automatically identified and discarded via the clean_rawdata procedure implemented in EEGLAB. The EEG was then cut into epochs (100 ms before stimulus onset to 600 ms after stimulus onset; 1 s padding was kept on both ends of each epoch to mitigate potential boundary artifacts in subsequent analyses). Epochs were z-score transformed relative to the baseline (-100-0 ms).

**ERP analysis**

Epochs were Gaussian-smoothed and averaged across trials to obtain the ERPs.

**ERSP analysis**

Before ERSP computation, the time-domain averaged response (i.e., the average of epochs) was subtracted from the epoch to assess the induced activity [34]. For each of the frequency bands (theta:2 to 7 Hz, alpha: 8 to 12 Hz, beta: 13 to 30 Hz, low-gamma: 31 to 60 Hz, and high-gamma: 61 to 150 Hz), epochs were band-pass filtered according to the specified frequency range, followed by calculation of magnitude of the Hilbert transform for each trial [33,40]. The epochs were then z-score transformed relative to the baseline. After that, same as ERP analysis, the processed epochs were Gaussian-smoothed and averaged across trials to obtain the ERSPs.

**Evaluation of responsive contacts**

For each EEG activity type (ERP or ERSP), an evaluation procedure was conducted to identify contacts where valid responses were evoked by the task stimuli [41]. For each contact, trials of expression conditions were averaged and a Wilcoxon signed-rank test was performed comparing the value of each time point in the post-stimulus duration (0-600 ms) against the baseline mean value (-100-0 ms). A contact was indicated as responsive if at least 10 (corresponding to 20 ms duration given the 500 Hz sampling rate) consecutive significant time points were identified (Wilcoxon signed-rank test $p<.001$; which passed Bonferroni correction). Response waveform quality was further verified via visual inspections. Only responsive contacts were subjected to subsequent analyses.

**Evaluation of important temporal periods**

The importance of temporal features in discriminating between facial expressions was evaluated by a PCA-based procedure [42,43]. For the post-stimulus timecourse of EEG activity (ERP or ERSP) at each contact, PCA was performed on a trial-averaged matrix with dimensionality T x C (where T referred to the number of time points and C referred to the number of expressions), with time points as features. PCs explaining 80% of the variance were extracted. The importance of the original features (i.e., time points) was then assessed via the loadings of the extracted PCs as sum(coef(:,1:k).^2*explained(1:k)) (where k denoted the number of PCs and coef denoted the loading coefficient). A higher importance score for a given time point indicated greater relative importance for discriminating between facial expressions. At least 10 (i.e., 20 ms length) consecutive important time points were considered. Important time periods were parametrically assessed according to the top 75%, top 50%, or top 25% of the total importance (i.e., the cumulative value of the temporal importance distribution). For cases with multiple discrete periods, instead of concatenating the non-contiguous segment periods, a single continuous period was used --- from the start of the first period to the end of the last period. This was more appropriate for catching the continuous temporal patterns in multivariate analysis.

**Pattern analysis**

For the EEG activity (ERP or ERSP) at each contact, the category specificity of the timecourse pattern was analyzed by pairwise comparisons between within-category and between-category similarities. Similarity between two signals x and y was estimated using a normalized Euclidean distance metric:

1-norm(x-y)/(norm(x)+norm(y)) (where x and y denoted two timecourses). EEG activity was computed separately for odd and even trials. For a given expression category, e.g., disgust, within-category similarity referred to the similarity between the activity temporal patterns of odd and even trials for disgust, i.e., disgust-disgust similarity. Between-category similarity referred to the similarity between the activity temporal pattern of odd trials for disgust and the activity temporal pattern of even trials for one of the other expressions. There were six between-category similarities: disgust-fear, disgust-anger, disgust-sadness, disgust-surprise, disgust-neutral, and disgust-happiness. The within-category similarity was compared against each between-category similarity, and a comparison was counted as correct if the within-category similarity value was greater than the between-category similarity value. The identification accuracy for disgust, was calculated as the proportion of correct comparisons. The statistical significance of an identification was evaluated using a random permutation procedure [42,44]. The identification process was repeated 200 times; in each iteration, trials were shuffled prior to implementing the pairwise comparison algorithm. A Wilcoxon signed-rank test was then used to determine whether the mean accuracy across permutations exceeded chance (50%) [24], and *p*<.05 after Bonferroni correction (for the number of between-category similarities) was considered as significant.

**Clustering analysis**

ERP timecourses of expressions at contacts with successful identification were subjected to a k-medoids clustering algorithm that partitions data points (i.e., the timecourses in the present study) into k groups. The algorithm minimizes the Euclidean distances between the cluster center (medoid: an actual data point within the cluster) and the points in each cluster. k-medoids clustering was selected over k-means clustering (in which the cluster center, centroid, is the mean of data points in a cluster) for two advantages: (1) k-medoids clustering exhibits greater robustness to outliers; (2) k-medoids clustering is better suited for scenarios requiring clarification of how well cluster means align with actual data points --- particularly when cluster means lack a clear, interpretable biological or methodological meaning. The variance-based elbow method was used to evaluate the optimal number of clusters for k between 2 and 10. Potential optimal k was considered as the value above which the reduction in explained variance (sum of distances between the data points and the cluster center, normalized to the sum for k=1) became more moderate [20]. The present study adopted the similarity-based approach given its simplicity, high efficiency, and particularly favorable interpretability (Haxby, 2012; Grootswagers et al., 2017).

## Acknowledgments

This work was supported by National Natural Science Foundation of China (31971033). The study was performed via the research platform "11buddy & I" (https://github.com/rwfwuwx/11buddy-and-I).

## Author contributions

**Yingyu Huang**: Software, Formal analysis, Writing - Original Draft, Writing - Review & Editing. **Lisen Sui:** Methodology, Investigation, Writing - Review & Editing, Supervision. **Liying Zhan**: Methodology, Software, Investigation, Formal analysis. **Chaolun Wang**: Methodology, Investigation, Formal analysis. **Zhihan Guo**: Methodology, Software, Investigation, Formal analysis. Yanjuan Li, **X X**: Methodology, Investigation. **Xiang Wu**: Conceptualization, Methodology, Software, Formal analysis, Writing - Original Draft, Writing - Review & Editing, Supervision,

Funding acquisition.

**Conflict of interest**

The authors declare no conflict of interest.

**Data availability Statement**

The data generated during and/or analyzed during the current study are available from the corresponding author on reasonable request.

# References

1. Oaten, M., Stevenson, R. J. & Case, T. I. Disgust as a disease-avoidance mechanism. *Psychological Bulletin* **135**, 303–321 (2009).
2. Kalat, J. W. *Biological Psychology*. (Cengage Learning, 2015).
3. Lamm, C. & Singer, T. The role of anterior insular cortex in social emotions. *Brain Struct Funct* **214**, 579–591 (2010).
4. Chapman, H. A. & Anderson, A. K. Understanding disgust. *Annals of the New York Academy of Sciences* **1251**, 62–76 (2012).
5. Gasquoine, P. G. Contributions of the Insula to Cognition and Emotion. *Neuropsychol Rev* **24**, 77–87 (2014).
6. Guillory, S. A. & Bujarski, K. A. Exploring emotions using invasive methods: review of 60 years of human intracranial electrophysiology. *Social Cognitive and Affective Neuroscience* **9**, 1880–1889 (2014).
7. Benarroch, E. E. Insular cortex: Functional complexity and clinical correlations. *Neurology* **93**, 932–938 (2019).
8. Soyman, E. *et al.* Intracranial human recordings reveal association between neural activity and perceived intensity for the pain of others in the insula. *eLife* **11**, e75197 (2022).
9. Schienle, A. *et al.* The insula is not specifically involved in disgust processing: an fMRI study. *NeuroReport* **13**, 2023 (2002).

10. Phan, K. L., Wager, T., Taylor, S. F. & Liberzon, I. Functional Neuroanatomy of Emotion: A Meta-Analysis of Emotion Activation Studies in PET and fMRI. *NeuroImage* **16**, 331–348 (2002).

11. Mazzola, L., Isnard, J., Peyron, R., Guénot, M. & Mauguière, F. Somatotopic organization of pain responses to direct electrical stimulation of the human insular cortex. *PAIN* **146**, 99 (2009).

12. Calder, A. J., Lawrence, A. D. & Young, A. W. Neuropsychology of fear and loathing. *Nat Rev Neurosci* **2**, 352–363 (2001).

13. Papagno, C. *et al.* Specific disgust processing in the left insula: New evidence from direct electrical stimulation. *Neuropsychologia* **84**, 29–35 (2016).

14. Sprengelmeyer, R. & Jentzsch, I. Event related potentials and the perception of intensity in facial expressions. *Neuropsychologia* **44**, 2899–2906 (2006).

15. Kurth, F., Zilles, K., Fox, P. T., Laird, A. R. & Eickhoff, S. B. A link between the systems: functional differentiation and integration within the human insula revealed by meta-analysis. *Brain Struct Funct* **214**, 519–534 (2010).

16. Deen, B., Pitskel, N. B. & Pelphrey, K. A. Three Systems of Insular Functional Connectivity Identified with Cluster Analysis. *Cerebral Cortex* **21**, 1498–1506 (2011).

17. Centanni, S. W., Janes, A. C., Haggerty, D. L., Atwood, B. & Hopf, F. W. Better living through understanding the insula: Why subregions can make all the difference. *Neuropharmacology* **198**, 108765 (2021).

18. Mukamel, R. & Fried, I. Human intracranial recordings and cognitive neuroscience. *Annu Rev Psychol* **63**, 511–537 (2012).

19. Mercier, M. R. *et al.* Advances in human intracranial electroencephalography research, guidelines and good practices. *NeuroImage* **260**, 119438 (2022).

20. Regev, T. I. *et al.* Neural populations in the language network differ in the size of their temporal receptive windows. *Nat Hum Behav* 1–19 (2024).

21. Fang, Z. *et al.* Human high-order thalamic nuclei gate conscious perception through the thalamofrontal loop. *Science* **388**, eadr3675 (2025).

22. Krolak-Salmon, P. *et al.* An attention modulated response to disgust in human

ventral anterior insula. *Annals of Neurology* **53**, 446–453 (2003).

23. Zhang, Y., Zhou, W., Huang, J., Hong, B. & Wang, X. Neural correlates of perceived emotions in human insula and amygdala for auditory emotion recognition. *NeuroImage* **260**, 119502 (2022).

24. Haxby, J. V. *et al.* Distributed and Overlapping Representations of Faces and Objects in Ventral Temporal Cortex. *Science* **293**, 2425–2430 (2001).

25. Grootswagers, T., Wardle, S. G. & Carlson, T. A. Decoding Dynamic Brain Patterns from Evoked Responses: A Tutorial on Multivariate Pattern Analysis Applied to Time Series Neuroimaging Data. *J Cogn Neurosci* **29**, 677–697 (2017).

26. Lundqvist, D., Flykt, A., & Öhman, A. The Karolinska Directed Emotional Faces – KDEF, CD ROM from Department of Clinical Neuroscience, Psychology section, Karolinska Institutet, ISBN 91-630-7164-9. (1998).

27. Goeleven, E., De Raedt ,Rudi, Leyman ,Lemke & and Verschuere, B. The Karolinska Directed Emotional Faces: A validation study. *Cognition and Emotion* **22**, 1094–1118 (2008).

28. Quabs, J. *et al.* Cytoarchitecture, probability maps and segregation of the human insula. *NeuroImage* **260**, 119453 (2022).

29. Kanwisher, N., McDermott, J. & Chun, M. M. The fusiform face area: A module in human extrastriate cortex specialized for face perception. *Journal of Neuroscience* **17**, 4302–4311 (1997).

30. Ganel, T., Valyear, K. F., Goshen-Gottstein, Y. & Goodale, M. A. The involvement of the "fusiform face area" in processing facial expression. *Neuropsychologia* **43**, 1645–1654 (2005).

31. Harry, B. B., Williams, M., Davis, C. & Kim, J. Emotional expressions evoke a differential response in the fusiform face area. *Front. Hum. Neurosci.* **7**, 3389 (2013).

32. Liu, M., Liu, C. H., Zheng, S., Zhao, K. & Fu, X. Reexamining the neural network involved in perception of facial expression: A meta-analysis. *Neuroscience & Biobehavioral Reviews* **131**, 179–191 (2021).

33. Pfurtscheller, G. & da Silva, F. H. L. Event-related EEG/MEG synchronization and desynchronization: basic principles. *Clinical Neurophysiology* **110**, 1842–1857 (1999).

34. Tallon-Baudry, C. & Bertrand, O. Oscillatory gamma activity in humans and its role in object representation. *Trends in Cognitive Sciences* **3**, 151–162 (1999).

35. Huang, W. *et al.* Direct interactions between the human insula and hippocampus during memory encoding. *Nat Neurosci* 1–9 (2025).

36. Willett, F. R. *et al.* A high-performance speech neuroprosthesis. *Nature* **620**, 1031–1036 (2023).

37. Gogolla, N. The insular cortex. *Current Biology* **27**, R580–R586 (2017).

38. Wu, X. & Zhang, D. Early induced beta/gamma activity during illusory contour perception. *Neurosci Letters* **462**, 244–247 (2009).

39. Meisler, S. L., Kahana, M. J. & Ezzyat, Y. Does data cleaning improve brain state classification? *Journal of Neuroscience Methods* **328**, 108421 (2019).

40. Baker, A. P. *et al.* Fast transient networks in spontaneous human brain activity. *eLife* **3**, e01867–e01867 (2014).

41. Metzger, S. L. *et al.* A high-performance neuroprosthesis for speech decoding and avatar control. *Nature* **620**, 1037–1046 (2023).

42. Willett, F. R., Avansino, D. T., Hochberg, L. R., Henderson, J. M. & Shenoy, K. V. High-performance brain-to-text communication via handwriting. *Nature* **593**, 249–254 (2021).

43. Metzger, S. L. *et al.* Generalizable spelling using a speech neuroprosthesis in an individual with severe limb and vocal paralysis. *Nat Commun* **13**, 6510 (2022).

44. Kochenderfer, M. J. & Wheeler, T. A. *Algorithms for Optimization*. (The MIT Press, Cambridge, Massachusetts, 2019).

# Supplementary materials

| Subjects | Age | Gender | Insula_L | Insula_R |
|---|---|---|---|---|
| S01 | 27 | M | - | 13 |
| S02 | 28 | M | - | 5 |
| S03 | 25 | M | 9 | - |
| S04 | 16 | M | - | 23 |
| S05 | 18 | F | - | 7 |
| S06 | 20 | M | 21 | - |
| S07 | 32 | M | - | 7 |
| S08 | 14 | M | - | 4 |
| S09 | 21 | F | 11 | - |
| S10 | 27 | F | - | 17 |
| S11 | 16 | F | 5 | - |
| Total | | | 46 | 76 |

**Table. S1. Information of insular contacts.** The number of insular contacts varied across subjects. For individual subjects, the contacts were largely implanted in a single hemisphere, with a greater total number of contacts within the right insula. (Two additional subjects were excluded from analyses due to poor behavioral performance. See Results, Behavioral performance).

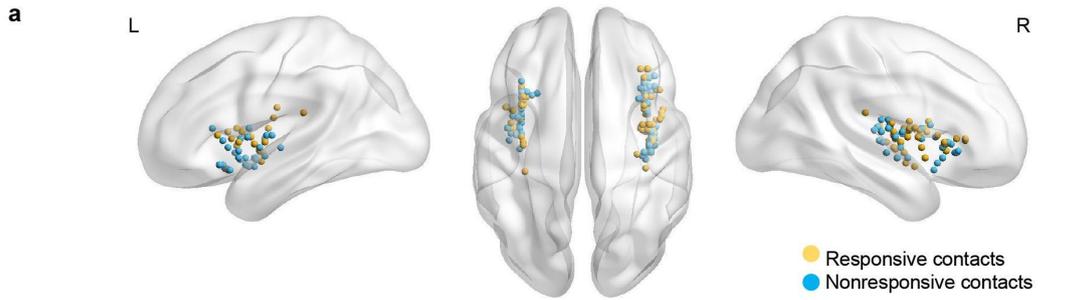

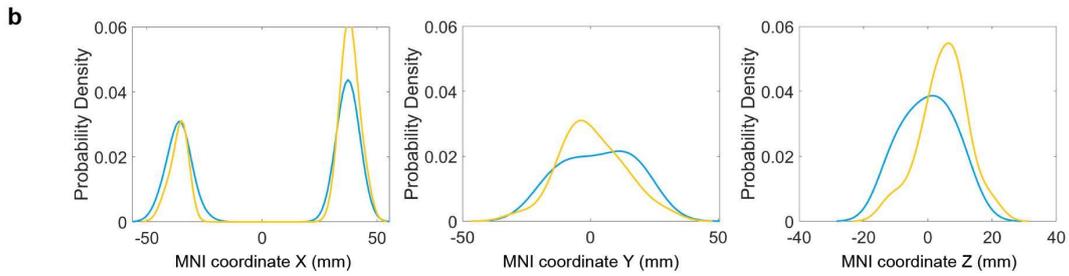

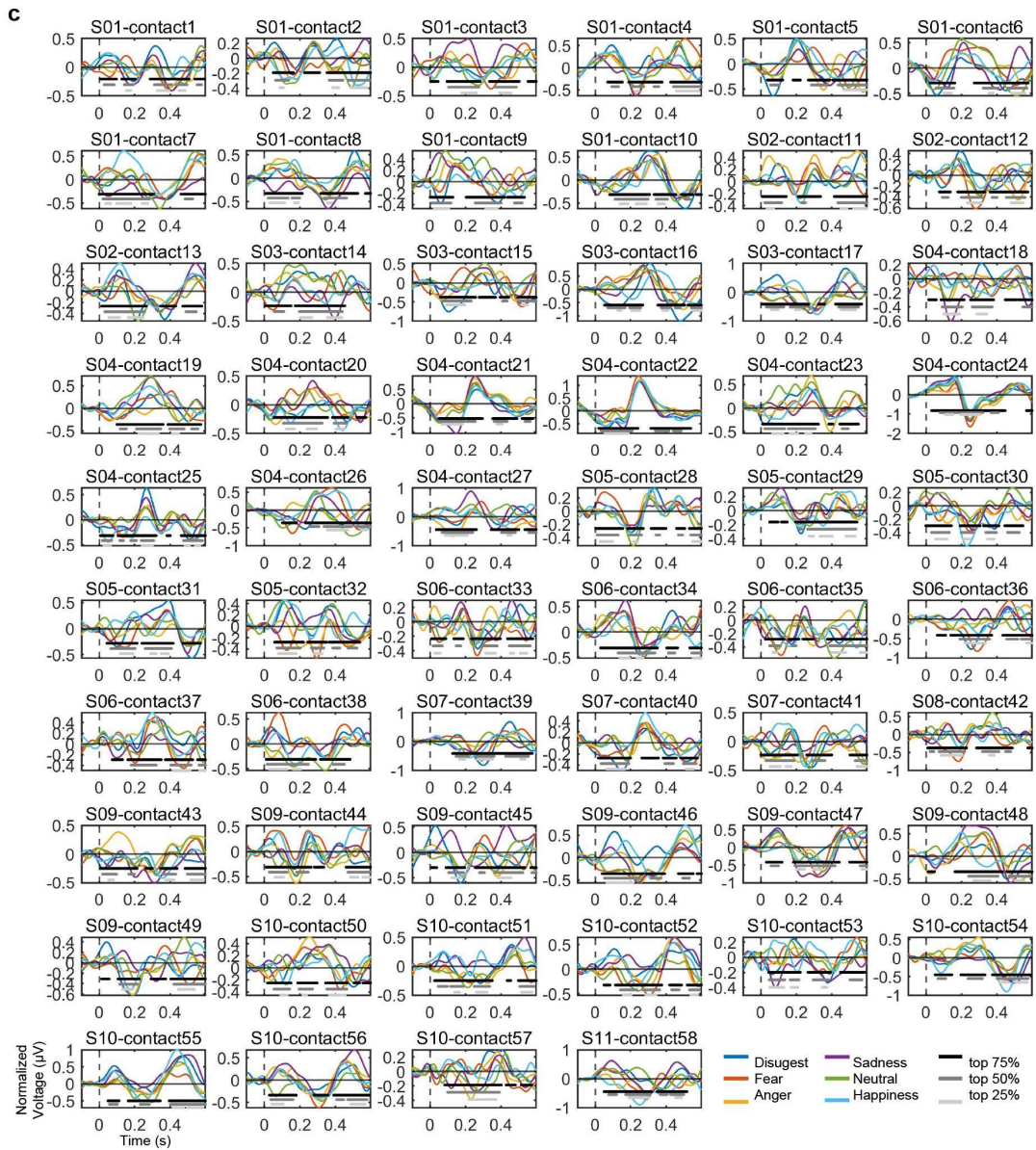

**Fig. S1. Illustration of responsive contacts and important time periods for insular ERPs**. **a**. Displays the task responsive insular contacts (yellow; n=58, left hemisphere 18, right hemisphere 40) and non-responsive insular contacts (blue; n=64; left hemisphere 28, right hemisphere 36) on a surface brain template. **b**. Distributions of responsive and non-responsive contacts, along the x, y, z coordinates, respectively. Responsive contacts were diffusely distributed across the insula (i.e., not concentrated in limited regions), and responsive and non-responsive insular contacts were intermixed except along the z axis; Kruskal-Wallis H test: x coordinate, $p>.05$; y coordinate, $p>.05$; z coordinate, $p=.001$. **c**. ERP timecourses of expressions at responsive contacts, along with the evaluated important time periods according to the top 75%, top 50%, and top 25% of total importance, respectively. Subject IDs and Contact IDs are annotated; Contact IDs correspond to responsive contacts only. Other conventions are as in Fig. 1-3.

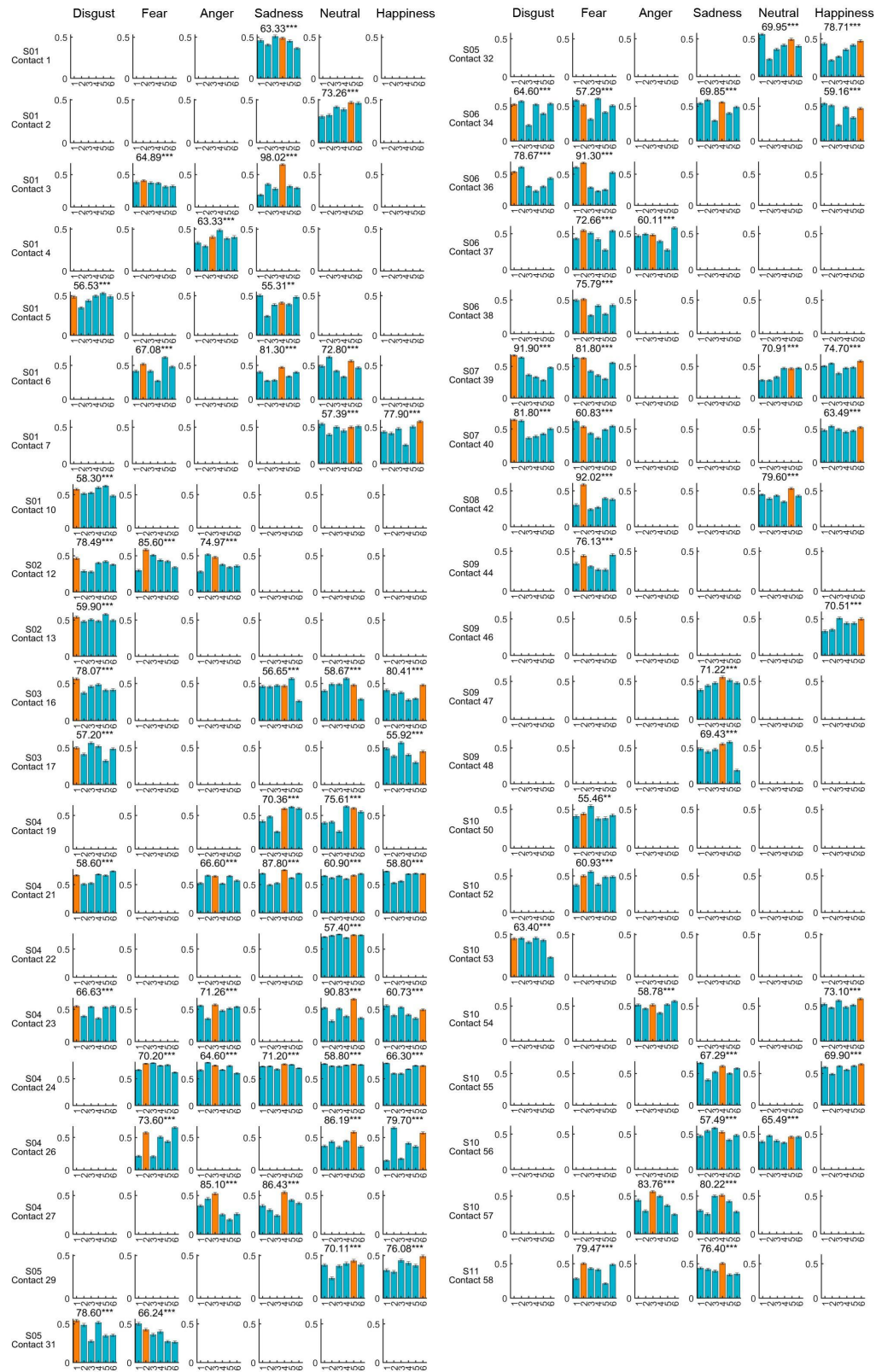

**Fig. S2. Illustration of within-category and between-category ERP pattern similarity values for insular contacts with successful identification**. Conventions are as in Fig. 2.

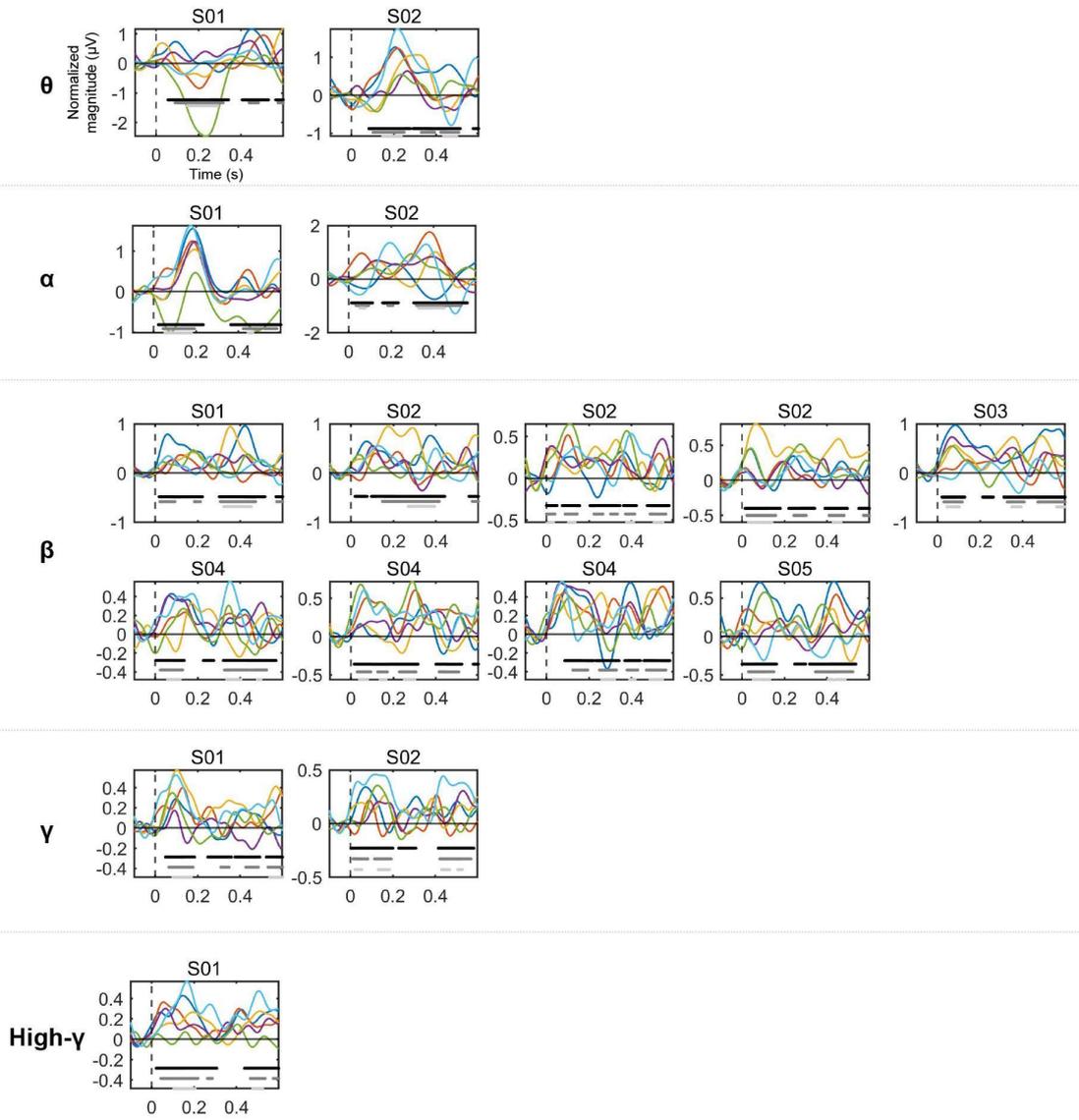

**Fig. S3. Illustration of responsive contacts and important time periods for insular ERSPs.** Conventions are as in Fig S1 and Fig. 5.

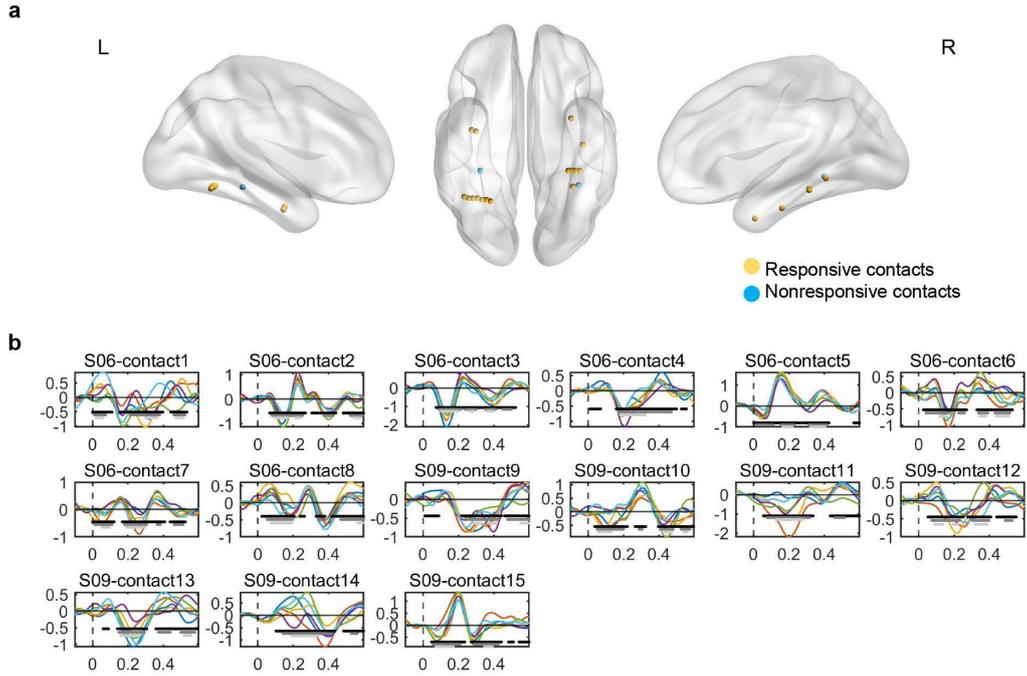

**Fig. S4. Illustration of responsive contacts and important time periods of FFA ERPs**. **a**. Displays the responsive FFA contacts (yellow; n=15, left hemisphere 9, right hemisphere 6) and non-responsive FFA contacts (blue; n=2; left hemisphere 2). b. Shows the ERP timecourses at the responsive contacts, along with the evaluated important time periods. Other conventions are as in Fig. S1 and Fig. 6.

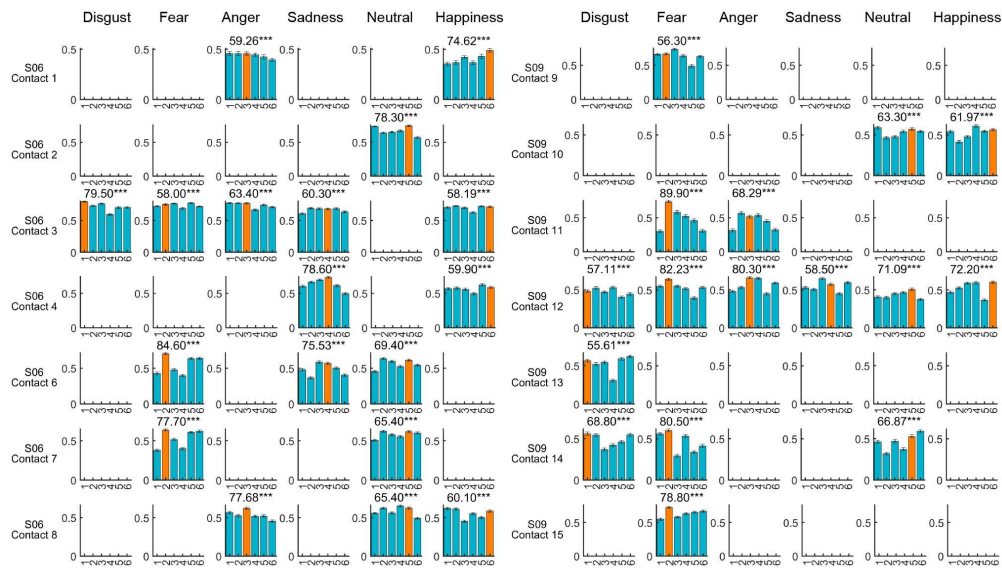

**Fig. S5. Illustration of within-category and between-category ERP pattern similarity values for FFA contacts with successful identification**. Conventions are as in Fig. S2 and Fig. 6.

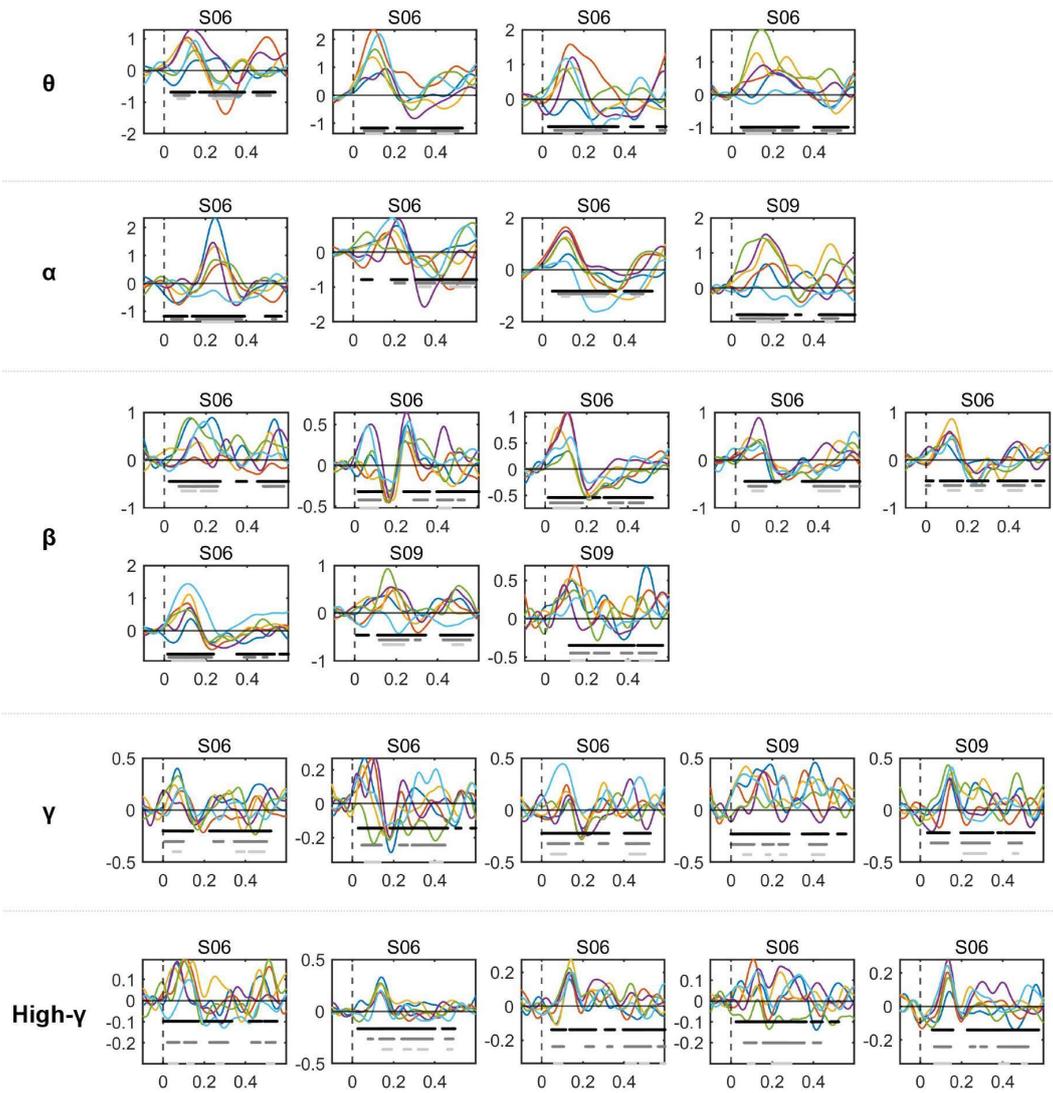

**Fig. S6. Illustration of responsive contacts and important time periods of FFA ERSPs.** Conventions are as in Fig S3 and Fig. 8.

**Expression identification by insular ERP patterns of the post-stimulus time duration**

Similar to the results based on the important time periods according to the top 50% of total importance (Fig. 2), the analysis based on the post-stimulus time period also yielded successful identification for all investigated expressions (Fig. S7). Specifically, disgusted, fearful, angry, sad, neutral, and happy faces were identified at 13, 16, 9, 17, 12, and 14 contacts, with corresponding mean±SD accuracies of 69.1±10.6, 69.2±9.0, 68.4±8.4, 70.0±11.2, 68.6±11.5, and 67.9±9.4, respectively. Overall, 41 among the 58 responsive contacts showed successful expression identification, yielding a mean±SD accuracy of 69.0±9.9 across all expressions.

In total, there were 86 successful identifications (one identification represented an expression being identified at a contact) for the top 50% important time periods (Fig. 2b), and 81 successful identifications for the post-stimulus time period (Fig. S7b), respectively. These two results exhibited substantial overlap, with 65 identifications shared in common. The mean±SD accuracy across all successful identifications was 70.8±10.4 for the top 50% important time periods and 69.0±9.9 for the post-stimulus time period, with no significant difference observed (Wilcoxon signed-rank test, $p>.05$). For the 71 overlapping successful identification, the mean±SD accuracy was 71.6±10.2 for the top 50% important time periods and 69.4±9.9 for the post-stimulus time period; also with no significant difference observed (Wilcoxon signed-rank test, $p>.05$).

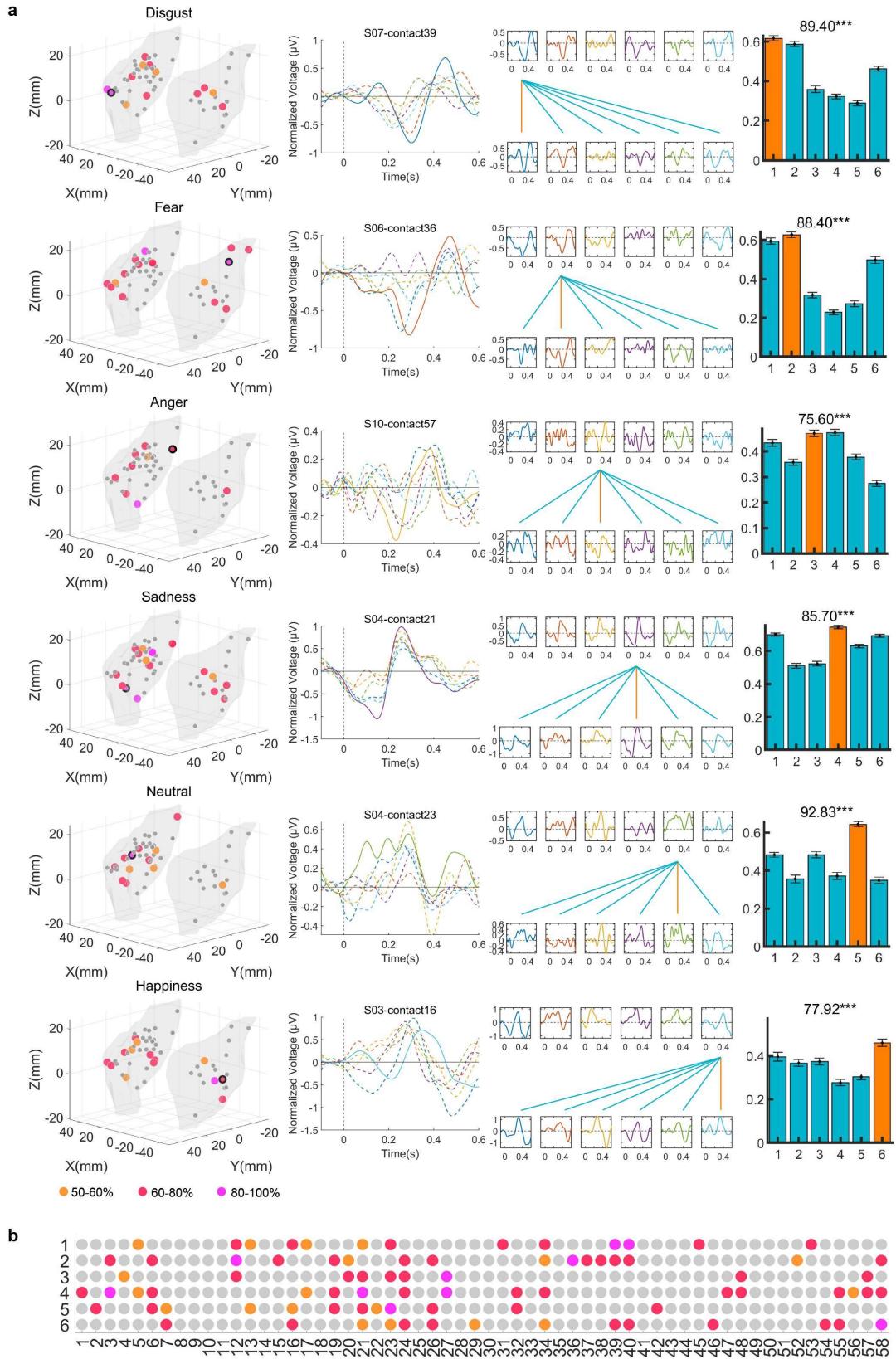

**Fig. S7. Successful facial expression identification by insular ERP patterns of the post-stimulus duration**. **a**. Shows the identification results, with one row per expression. **b**. Matrix representation of successful identifications. These results were based on the post-stimulus time duration (i.e., all post-stimulus time points corresponding to 100% importance), instead of the important time periods according to the top 50% of total importance (Fig. 2). Other conventions are as in Fig. 2.

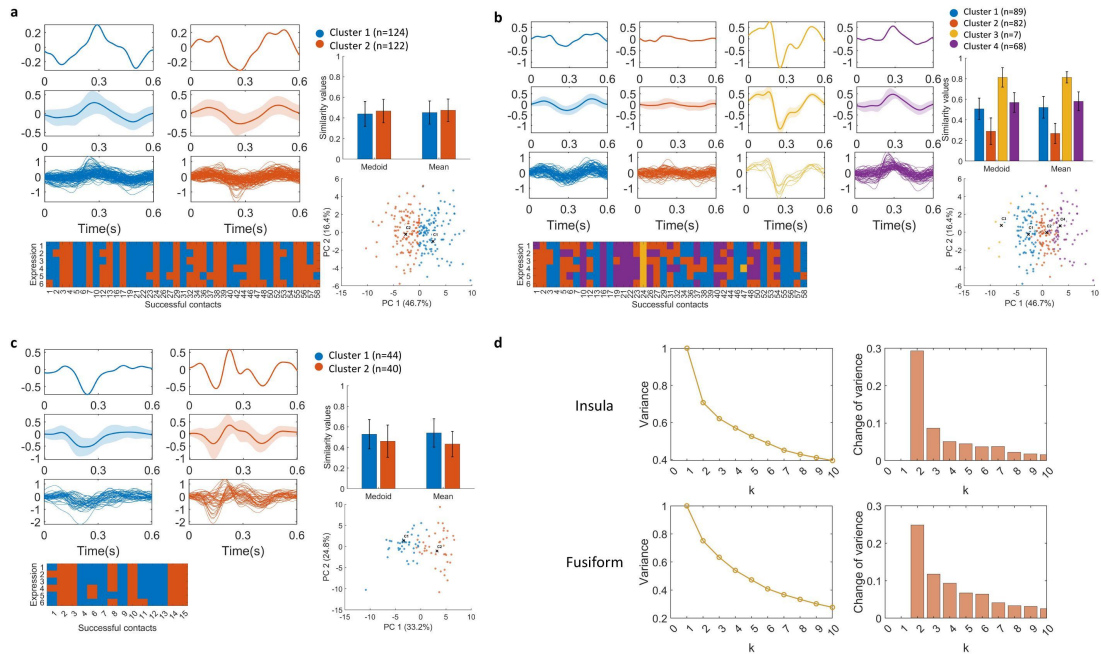

**Fig. S8. Supplementary clustering analysis. a**. Insular clustering results for k = 2. **b**. For insular data, no discernible drop in variance change was observed at k = 4. Here, we further tested insular clustering results for k = 4, as a control to the k=4 analysis of FFA data (see Fig. 7). Results of k=4 were similar to results of k=3 (see Fig. 4), except that Cluster 2 for k=3 that showed a negative deflection was further partitioned into two clusters with negative deflection profiles (Clusters 1 and 3 for k=4). In other words, insular timecourses did not cluster adequately by increasing k, particularly for the subset of timecourses (in Cluster 3 at k=3 or in Cluster 2 at k=4) that did not exhibit consistent response patterns. **c**. FFA clustering results for k = 2. Timecourses were not clustered adequately. **d**. Displays, for the insula (top) and FFA (bottom), the variance (left) and variance change (right) as a function of k, calculated by the k-means method; these results were comparable to those calculated using the k-medoids method (see Fig. 4 and 7) results. Other conventions are as in Fig. 4 and 7.